\documentclass[aps,pra,reprint,superscriptaddress]{revtex4-2}

\usepackage{amsmath,amssymb,bm,physics}
\usepackage{graphicx}
\usepackage{hyperref}
\usepackage{float}
\usepackage[caption=false]{subfig}
\usepackage{comment}
\usepackage{verbatim}

\hypersetup{
    colorlinks=true,
    linkcolor=black,    % Color of internal links (e.g., reference numbers)
    citecolor=blue,    % Color of citation links (e.g., \cite{})
    urlcolor=blue       % Color of URL links (e.g., \url{})
}

\newcommand{\CZthree}{\mathrm{CZ}_3}
\newcommand{\avgF}{F_{\mathrm{avg}}}

\begin{document}

\title{Reinforcement Learning for Robust Calibration of Multi-Qudit Quantum Gates}

\author{Amine Jaouadi}
\email{ajaouadi@ece.fr}
\affiliation{LyRIDS, ECE Engineering School, OMNES Education, 10 rue Sextius Michel, 75015, Paris, France}
\author{Sahel Ashhab}
%\email{ashhab@nict.go.jp}
\affiliation{Advanced ICT Research Institute, National Institute of Information and Communications Technology,
4-2-1, Nukui-Kitamachi, Koganei, Tokyo 184-8795, Japan}
\affiliation{Research Institute for Science and Technology, Tokyo University of Science, 1-3 Kagurazaka, Shinjuku-ku, Tokyo 162-8601, Japan}

\date{\today}

% ------------------------------------------------------
% Abstract
% -----------------------------------------------------
\begin{abstract}
Higher-dimensional quantum systems, such as qudits, offer
architectural and algorithmic advantages over qubits, but their
increased spectral crowding and limited controllability render
high-fidelity quantum gates particularly challenging. We propose a
hybrid optimization framework that integrates optimal control theory
methods with contextual deep reinforcement learning to achieve robust
controlled-phase gates on two qutrits. Optimal control is first used
to design high-fidelity control pulses for a nominal system model.
Reinforcement learning is then employed as a calibration stage that learns small residual corrections to these pulses in the presence of static model mismatch, thereby preserving good gate performance under parameter uncertainties and/or variations. By learning structured,
low-dimensional residual corrections conditioned on device-specific
parameter variations, reinforcement learning enhances the transfer robustness of nominally optimal but parameter-sensitive control solutions across ensembles of devices. Crucially, the reinforcement
learning step in our framework does not compete with the optimal
control step but provides the adaptability required for realistic
hardware, systematically reducing the sensitivity to parameter
fluctuations. Our results establish reinforcement learning as a
practical and scalable ingredient for robust calibration of quantum gates in high-dimensional systems. 

\end{abstract}

\maketitle

% ======================================================
% I. INTRODUCTION
% ======================================================
\section{Introduction}
\label{sec:intro}

Qudits---$d$-level quantum systems with $d>2$---are increasingly recognized as promising building blocks for quantum information processing, offering larger local Hilbert spaces, richer entangling operations, and potential resource savings in circuit depth and error correction~\cite{Bullock2005,Lanyon2009,Chi2022}. In particular, qutrits ($d=3$) have recently been explored in several hardware platforms, including trapped ions and superconducting circuits~\cite{Bianchetti2010,Blok2021,Roy2023,Goss2022, Kononenko2021, Glaser2015, Morvan2021,Yurtalan2020, Ringbauer2022}. Superconducting transmons are especially attractive qutrit candidates because their weak anharmonicity naturally provides access to the first three energy levels with gate times comparable to qubit operations~\cite{Basyildiz2025}. It should be noted that other superconducting qubit designs, such as the capacitively-shunted flux qubit, also allow access to higher energy levels~\cite{You2007}.

Recent experiments have demonstrated high-fidelity single-qutrit control, randomized benchmarking, and two-qutrit entangling gates in transmon devices~\cite{Goss2022,Roy2023,Subramanian2023, Kononenko2021}. Nevertheless, engineering robust two-qutrit gates remains challenging: the richer level structure enhances controllability but also introduces additional leakage channels and strong sensitivity to device parameters such as transition frequencies and coupling strengths~\cite{Goss2022,Subramanian2023}. Fabrication variability and slow drifts in these parameters further accentuate the need for automated calibration strategies beyond manual tune-up.

Quantum optimal control theory (QOCT), and in particular gradient ascent pulse engineering (GRAPE)~\cite{Heeres2017, Poggi2024, Jaouadi2013, Vranckx2013}, has emerged as a key tool for numerically designing high-fidelity gates in effective models, such as superconducting circuits~\cite{Khaneja2005,Koch2022,Egger2013,Kelly2014}. GRAPE efficiently computes gradients of the fidelity with respect to the control waveform and has produced fast, high-fidelity two-qubit gates~\cite{Egger2013, Ashhab2012, Ghosh2013}, leakage-suppressed single-qubit and two-qubit gates~\cite{Motzoi2009, Ashhab2022} and few-qudit gates~\cite{Basyildiz2025,Ashhab2026}. However, the performance of QOCT-optimized control pulses is fundamentally tied to model accuracy: deviations of the true device Hamiltonian from the nominal design, whether from fabrication spread or temporal drift, can degrade gate fidelity~\cite{Koch2022,Egger2014}. Accordingly, robustness has become a key objective in multi-level superconducting quantum control~\cite{Poggi2024}.

In parallel, deep reinforcement learning (DRL) has shown promise as a largely \emph{model-free} approach to quantum control, enabling the direct optimization of control pulses through reward feedback rather than explicit system modeling~\cite{Bukov2018,Niu2019,Liu2025,Jaouadi2024}. These methods have been successfully implemented using algorithms such as Soft Actor–Critic (SAC)~\cite{Haarnoja2018}, Twin Delayed Deep Deterministic Policy Gradient (TD3)~\cite{Fujimoto2018}, Deep Deterministic Policy Gradient (DDPG)~\cite{Lillicrap2015}, and Proximal Policy Optimization (PPO)~\cite{Schulman2017}. Yet recent studies have highlighted intrinsic limitations of \emph{model-free DRL} when applied directly to high-dimensional pulse synthesis. In particular, while model-free methods can be convenient because of their versatility and minimal requirements, they do not benefit from physics-informed rules that conventional QOCT techniques rely on. More specifically,the action space in a quantum control problem scales with the number of pulse segments, often reaching hundreds or thousands of degrees of freedom. A recent study \cite{Li2025_DRLfD} showed that standard DRL agents struggle to discover high-fidelity control pulses when trained from scratch in high-dimensional control spaces. In such settings, DRL stagnates at low reward levels and exhibits poor convergence unless supplemented with prior knowledge or demonstrations. As will be demonstrated below, these challenges are indeed present in our study, where the enlarged Hilbert space and increased control complexity substantially expand the optimization landscape, making purely model-free DRL increasingly difficult to train efficiently and preventing the algorithms considered here from discovering high-fidelity solutions within practical training budgets.

Complementary insights into the control of multi-level superconducting
circuits were recently provided in ~\cite{Basyildiz2025,Ashhab2026}, in which the authors analyzed fundamental speed limits for a two-qutrit controlled-phase
gate implemented with superconducting qudits.
Using gradient-based optimal control to construct time-optimal protocols
together with analytical quantum speed limit arguments. They showed that
even for idealized models the achievable gate performance is constrained
by the Hamiltonian structure and control resources.
On the other hand, these results emphasize that, on a nominal device, gradient-based QOCT
can already saturate the accessible control landscape for two-qutrit gates, leaving little room for improvement by unguided learning-based approaches.
These observations suggest that DRL is unlikely to replace QOCT for pulse synthesis on nominal, well-characterized devices, but may instead play a complementary role in addressing model mismatch
and calibration challenges. From a practical perspective, repeated re-optimization of QOCT pulses for each device instance or parameter drift can be computationally demanding and difficult
to scale. This motivates approaches in which a high-quality nominal pulse is computed once, and subsequent corrections are obtained through lightweight calibration procedures rather than repeated high-dimensional optimization.

\vspace{0.5cm}

Motivated by these challenges, we introduce a hybrid QOCT+DRL framework in which
QOCT and DRL play complementary roles: QOCT provides
an open-loop strategy to obtain a nominal high-fidelity pulse, while DRL provides a closed-loop calibration stage that
learns residual corrections and tweaks the pulse to mitigate fidelity loss caused by model mismatch between the nominal and actual device parameters.

The novelty of our contribution can be summarized as follows:

\begin{itemize}
    \item \emph{DRL on top of QOCT pulses.} Rather than asking DRL to discover an optimal pulse in a high-dimensional space (a task known to be prohibitive~\cite{Li2025_DRLfD}), we compute a pair of nominal GRAPE pulses, computed once on the nominal Hamiltonian for a two-qutrit $\CZthree$ gate and train DRL agents to output \emph{low-dimensional, smooth residual corrections}. This approach bypasses the failure mode of DRL in large action spaces.
    \item \emph{Contextual bandit formulation of device-aware calibration.} Each episode corresponds to a different noisy device instance, and the observation encodes normalized parameter deviations. The reward is the \emph{incremental fidelity gain over the QOCT baseline}, forcing the agent to learn how to achieve robustness rather than re-learn the gate. 
    \item \emph{Cosine-basis parametrization.} Residual corrections are parametrized in a truncated discrete cosine basis, drastically reducing the action dimensionality and enforcing smoothness.
    \item \emph{Comparison of multiple DRL algorithms.} SAC, TD3, DDPG, and PPO are evaluated under identical conditions. Importantly, \emph{none} of the agents outperform QOCT on the nominal Hamiltonian—consistent with the failure of DRL in high-dimensional unconstrained optimization~\cite{Li2025_DRLfD}. This observation further supports our use of DRL as a calibration (i.e.~fine-tuning) tool rather than a replacement for QOCT-based pulse design.

\end{itemize}

The paper is organized as follows. Section~\ref{sec:model} introduces the effective two-qutrit model, the target controlled-phase gate, and the fidelity measure, together with an overview of the control architecture. Section~\ref{sec:QOCT} presents the GRAPE-based optimal control procedure and demonstrates its convergence on the nominal device. Section~\ref{sec:DRL} describes the residual reinforcement-learning framework, including the noise model, cosine-basis parametrization, and contextual bandit formulation. Section~\ref{sec:DRL_Algo} summarizes the deep reinforcement-learning algorithms and the training protocol used in this work. Section~\ref{sec:results} reports the numerical results, beginning with an assessment of QOCT and DRL performance on the nominal Hamiltonian, and then addressing performance on a single noisy device, ensemble robustness, and the structure of the optimized pulses. Section~\ref{sec:discussion} discusses implications for superconducting-qutrit hardware and places the proposed approach in the broader context of hybrid open/closed-loop optimal control. Finally, Section~\ref{sec:conclusion} concludes the discussion
. Appendix~\ref{app:alt_gate} provides additional validation using an alternative two-qutrit controlled-phase gate.

% ======================================================
\section{Two-qutrit model, target gate, and fidelity}
\label{sec:model}

\subsection{Effective two-qutrit Hamiltonian}

We consider two coupled transmon-like qutrits with annihilation operators $a_1$ and $a_2$, truncated to the lowest three energy levels. The effective Hamiltonian in the lab frame reads
\begin{align}
H(t) &= H_0(\bm{\lambda}) + \sum_{i=1}^2 \epsilon_i(t) H_{c,i},
\label{eq:H_total}
\end{align}
with
\begin{align}
H_0(\bm{\lambda}) &= \sum_{i=1}^2 \left( \omega_i n_i + \chi_i a_i^{\dagger 2} a_i^{2} \right)
+ g \left( a_1 + a_1^\dagger \right)\left( a_2 + a_2^\dagger \right),
\label{eq:H0}\\
H_{c,1} &= a_1 + a_1^\dagger,\qquad
H_{c,2} = a_2 + a_2^\dagger,
\end{align}
where $n_i = a_i^\dagger a_i$, $\bm{\lambda} = (\omega_1,\omega_2,\chi_1,\chi_2,g)$ collects the device parameters, and $\epsilon_i(t)$ denote the classical drive envelopes.

In the numerical implementation, we represent $a_i$ as ladder operators acting on a three-dimensional Hilbert space. In the GRAPE algorithm, time evolution is computed using piecewise-constant controls over $N$ time slices of duration $\Delta t$, 
\begin{equation}
\epsilon_i(t) = \epsilon_i^{(j)},\qquad t \in [j\Delta t, (j+1)\Delta t),
\end{equation}
leading to the propagator
\begin{equation}
U[\bm{\epsilon};\bm{\lambda}] = \prod_{j=N-1}^{0} \exp\!\left[-i H^{(j)}(\bm{\lambda})\Delta t\right],
\end{equation}
with $H^{(j)}(\bm{\lambda}) = H_0(\bm{\lambda}) + \sum_i \epsilon_i^{(j)} H_{c,i}$.

\subsection{CZ$_3$ gate}

We target a two-qutrit controlled-phase gate $\CZthree$ acting on the computational basis
$\{ \ket{00},\ket{01},\ket{02},\ket{10},\dots,\ket{22} \}$ as
\begin{equation}
U_{\CZthree} =
\mathrm{diag}(1,1,1,\,
1,e^{2\pi i/3},e^{-2\pi i/3},\,
1,e^{-2\pi i/3},e^{2\pi i/3}),
\label{eq:CZ3}
\end{equation}
where the diagonal is written in lexicographic order of the two-qutrit basis. This gate is universal for ternary computation when combined with suitable local operations~\cite{Bullock2005,Subramanian2023} and closely related to the two-qutrit entangling gates realized experimentally in superconducting circuits~\cite{Goss2022,Roy2023}.

\subsection{Average gate fidelity}

We quantify performance using the average gate fidelity between the realized unitary $U$ and the target $U_{\CZthree}$,
\begin{equation}
\avgF(U,U_{\CZthree}) =
\frac{ \left| \Tr\!\left( U_{\CZthree}^\dagger U \right) \right|^2 + d^2 }{d^2(d+1)}.
\label{eq:Favg_def}
\end{equation}
This expression is derived from the general formula for the average gate fidelity of a unitary channel given in Ref~\cite{Nielsen2002}, specialized here to the case of a $d$-dimensional unitary operation.
 
In our simulations, $U$ is obtained either by propagating with QOCT-only pulses or with QOCT+DRL-corrected pulses on a specified device instance.

% ======================================================
\section{Nominal Device GRAPE optimization}
\label{sec:QOCT}

We first design nominal control pulses $(\epsilon_1^{\mathrm{QOCT}},\epsilon_2^{\mathrm{QOCT}})$ using GRAPE~\cite{Khaneja2005} as implemented in the QuTiP \texttt{pulseoptim} module~\cite{Johansson2013}. The drift Hamiltonian is defined at a nominal parameter point $\bm{\lambda}_0 = (\omega_{1,0},\omega_{2,0},\chi_{1,0},\chi_{2,0},g_0)$, selected to reflect typical parameters of a two-transmon device.

A useful reference timescale for entangling gates in superconducting circuits is obtained by truncating the system to two levels per qubit. In this case, the minimum time required to implement a CNOT or CZ gate under a coupling of strength $g$ is $T_0 = \frac{\pi}{4g}$.
More generally, quantum speed limits for two-body entangling gates scale as $T_{\min} \sim 1/g$, as discussed in Ref~\cite{Basyildiz2025}, with comparable prefactors for qubit and qutrit CZ gates. Accordingly, the characteristic timescale for implementing a two-qutrit CZ gate remains of the same order as in the qubit case, i.e., $T_{\min}^{(2\text{-qutrit})} \sim T_0 = \frac{\pi}{4g}$.

In our simulations, we use $g=0.0025$, for which
$T_0 \approx 314$ and $T_{\min}^{(2\text{-qutrit})}\approx 628$.
When we perform QOCT optimization with pulse times $T$ below this threshold, the optimization fails to reach high fidelity, as expected. In the remainder of this paper, we set the pulse time at $T=1600$, which places the system
well above the minimum-time regime and allows reliable convergence to unit fidelity.

The GRAPE cost function is the infidelity $1-F$, where $F$ is the average gate fidelity. The optimizer terminates when either the target infidelity threshold (here $\lesssim 10^{-10}$) or the maximum number of iterations is reached.

% ======================================================
\section{Noise model and residual reinforcement learning}
\label{sec:DRL}

\subsection{Device parameter noise}

In any physical implementation of qutrits, transition frequencies and interaction parameters will not be perfectly stationary and may vary for example because of fabrication variations and/or slow drift. Furthermore, one must also deal with the uncertainty caused by limited calibration accuracy and imperfect Hamiltonian modeling. We focus on superconducting qutrits as the specific technology of interest in this work. In superconducting processors, parameter variations between different devices during fabrication are unavoidable. These variations can be on the order of MHz or more \cite{Kreikebaum}. Temporal fluctuations in the parameters in a single device are strongly dependent on the exact circuit design. For example, designs involving loops, such as tunable-frequency qutrits and tunable couplers, tend to be significantly more prone to exhibit parameter fluctuations, because of fluctuating external magnetic fields. With this point in mind, short-time fluctuations of qubit frequencies range from sub-kHz to tens of kHz \cite{Agarwal2025,Burnett2019}. Strongly coupled two-level system (TLS) defects can cause fluctuations on the order of 1 MHz. Considering that qutrit transition frequencies are typically a few GHz, the relative variations are in the range $10^{-7}$- $10^{-3}$. There have not been systematic investigations presenting precise experimental characterization of temporal fluctuations in inter-qutrit coupling strengths in the literature. However, these are also expected to be below 1 kHz for fixed coupling designs but reach tens of kHz for tunable coupling designs. Coupling strengths are typically a few MHz, which means that the relative variations are in the range $10^{-4}$- $10^{-2}$. It should also be noted that the presence of loops used to achieve tunability of one or more parameters can naturally lead to correlations between the variations of qutrit frequencies and coupling strengths.

We take the highest numbers in the previous paragraph as representative of worst-case scenarios for the parameter fluctuations. We will see shortly that, with these numbers, fluctuations in the qutrit transition frequencies are more detrimental than fluctuations in the coupling strength. This result can be intuitively understood based on the fact that two-qutrit gates often rely on control pulses that are resonant with individual qutrit transition frequencies \cite{Paraoanu2006,Rigetti2010,DeGroot2010}. Variations in this parameter can therefore take a control pulse out of resonance and significantly alter the gate dynamics. In contrast, the coupling strength determines the speed of the entangling dynamics. Therefore, small fluctuations in the coupling strength will result in slightly undershooting or overshooting the target gate.  
For the fluctuation levels considered here, frequency uncertainties can significantly degrade gate performance, demonstrating that control pulses optimized for a nominal device may fail to achieve high fidelity when applied to devices with different parameter realizations. As shown below, contextual DRL effectively compensates for these variations by adapting the control pulses to the device-specific parameters, recovering most of the performance lost due to parameter uncertainty. Future experiments that require significantly lower infidelities than present-day ones can use the same approach to suppress even small gate infidelities resulting from small parameter variations.

Apart from the question of parameter variations, there is another one about the parameter uncertainty, i.e.~how accurately we can know the parameters. The parameters that enter in the theoretical model of a specific experiment are inferred from calibration measurements and are therefore subject to estimation uncertainty. The achievable estimation accuracy is limited by measurement noise and finite coherence times ($T_1, T_2 \sim 100~\mu\mathrm{s}$ to $1~\mathrm{ms}$), which constrain the precision of the extracted Hamiltonian parameters to the kHz level.

To model fabrication variability and slow drift, we introduce
fluctuations in the two qutrit frequencies
$\omega_1=\omega_{1,0}+\delta\omega_1$ and
$\omega_2=\omega_{2,0}+\delta\omega_2$ .
The frequency fluctuations considered here should be interpreted as effective residual parameter mismatches that remain after standard device calibration procedures, rather than the full fabrication-induced spread across nominally identical devices. Our objective is therefore to assess the ability of the DRL calibration layer to compensate realistic residual modeling and calibration errors around a nominally characterized device.
We will set the offsets $\delta\omega_j$ by drawing random numbers independently from Gaussian distributions 
$\delta\omega_1,\delta\omega_2 \sim \mathcal{N}(0,\sigma_\omega^2)$ with $\sigma_\omega = 10^{-3}$.
In line with the estimates given above, we keep the inter-qutrit coupling fixed at its
nominal value $g$. It is straightforward to generalize our analysis and include variations in $g$, if we deal with an experimental situation where these variations have a significant impact on the dynamics.

%\begin{figure}[h]
%    \centering
 %   \includegraphics[width=\columnwidth]{noise_distribution.png}
%    \caption{Noise statistics for the device parameters. Histograms of $10^5$ samples of $\delta\omega_1$, $\delta\omega_2$ (left and center) and coupling fluctuation $\delta g$ (right), generated from Gaussian distributions with standard deviations $\sigma_\omega = 10^{-3}$ and $\sigma_g = 5\times 10^{-5}$. \textcolor{red}{(The last histogram is broader than $10^{-5}$. This is probably a typo. But first we should discuss whether we will keep these distributions or use different ones.)} Dashed vertical lines indicate the mean (zero), dotted lines indicate $\pm 3\sigma$. These distributions define the noisy device ensemble used for the robustness study in Fig.~\ref{fig:ensemble_bar}.}
%    \label{fig:noise}
%\end{figure}

%To better understand the undeDRLying disorder, we visualize the noise model by sampling $10^5$ independent triples $(\delta\omega_1,\delta\omega_2,\delta g)$ from Gaussian distributions and building empirical histograms as shown in Fig.~\ref{fig:noise}. Each panel corresponds to one parameter and displays the probability density estimated from the samples; dashed vertical lines mark the mean (zero) and dotted lines mark the $\pm 3\sigma$ intervals. The large number of samples ensures that the depicted distributions faithfully represent the device ensemble seen by the DRL agents and used in the robustness evaluation.

\subsection{Cosine-basis residual parametrization}

DRL can be incorporated into quantum control in several ways. One possibility is to take a pulse as defined in the GRAPE algorithm and optimize all the pulse amplitudes using DRL, i.e.~treating the control pulse amplitude in each time segment as an action variable in the DRL calculation. Another approach is to update the control amplitudes sequentially during the evolution. While such formulations are conceptually straightforward, they lead to high-dimensional action spaces and often suffer from inefficient exploration in complex quantum-control landscapes.
In the present work, our objective is different. Rather than synthesizing control pulses from scratch, we seek to calibrate an already high-fidelity QOCT solution in the presence of device-dependent parameter mismatch. Consequently, the DRL agent is tasked only with learning smooth residual corrections to the nominal QOCT pulse. To achieve this, we represent the residual controls using a truncated cosine basis.
Let $\epsilon_i^{\mathrm{QOCT}}(t)$ denote the QOCT pulse applied to control channel $i$. The calibrated pulse is written as
\begin{equation}
\epsilon_i(t)
=
\epsilon_i^{\mathrm{QOCT}}(t)
+
\Delta\epsilon_i(t),
\end{equation}
where $\Delta\epsilon_i(t)$ is a residual correction generated by the DRL policy.
The control pulse that we calculate using the nominal device parameters is discretized into $N$ time segments. However, rather than introducing one correction parameter per segment, the correction is expanded on a truncated cosine basis. The $j$th step in the discretized pulse then acquires the correction
\begin{equation}
\Delta \epsilon_i^{(j)}
=
\sum_{k=1}^{K}
c_{i,k}
\cos\left[
\frac{\pi k (j+1/2)}{N}
\right],
\label{eq:cos_basis}
\end{equation}
where $j=0,\ldots,N-1$ labels the pulse segments, $k=1,\ldots,K$ labels the cosine basis functions retained in the truncated
expansion, and $c_{i,k}$ are the coefficients produced by the DRL agent.
This parametrization provides two important advantages. First, it substantially reduces the dimension of the optimization problem. Instead of modifying hundreds of pulse amplitudes independently, the agent only needs to determine a small set of coefficients. Second, the cosine basis naturally generates smooth corrections and helps regularize the learning process.
For the two control channels considered here, the DRL action consists of the set of coefficients
\begin{equation}
a=
{c_{1,1},\ldots,c_{1,K},
c_{2,1},\ldots,c_{2,K}},
\end{equation}
resulting in an action-space dimension of $2K$. In the simulations presented below, we use $K=20$, yielding a 40-dimensional action space. The resulting calibrated pulses are then evaluated on the noisy device and optimized through the DRL procedure described in the following subsection.

\subsection{Contextual bandit environment}

We formulate the calibration problem as a one-step contextual bandit. Each \emph{episode} proceeds as follows:

\begin{enumerate}
    \item Sample a device instance by drawing $(\delta\omega_1,\delta\omega_2)$ from the noise distribution.
    \item Construct the effective Hamiltonian $H_0(\bm{\lambda})$ with $\bm{\lambda}=(\omega_1,\omega_2,\chi_1,\chi_2,g)$.
    \item Compute the QOCT baseline fidelity for this device $F_{\mathrm{QOCT}}(\delta\omega_1,\delta\omega_2) =
    \avgF\big( U[\bm{\epsilon}^{\mathrm{QOCT}};\bm{\lambda}],\, U_{\CZthree}\big)$.
    \item Provide the agent with a normalized context vector
    $\bm{o} =
    \left(
    \frac{\delta\omega_1}{\sigma_\omega},
    \frac{\delta\omega_2}{\sigma_\omega}
    \right)$,
    which lies in a bounded subset of $\mathbb{R}^2$.
    \item The agent outputs an action $\bm{a}\in [-1,1]^{2K}$, corresponding to cosine coefficients for the two drives.
    \item Build residual pulses via Eq.~\eqref{eq:cos_basis}, form the total pulses $\bm{\epsilon}^{\mathrm{tot}}$, propagate the system, and compute $F_{\mathrm{DRL}}(\delta\omega_1,\delta\omega_2) =
    \avgF\big( U[\bm{\epsilon}^{\mathrm{tot}};\bm{\lambda}],\, U_{\CZthree}\big)$.
    \item Return a scalar reward $r = F_{\mathrm{DRL}} - F_{\mathrm{QOCT}}$
    and terminate the episode.
\end{enumerate}

By construction, $r>0$ if and only if the DRL-corrected pulses outperform the QOCT baseline on that particular device instance. This reward shaping makes the learning problem explicitly \emph{residual}: the agent is incentivized to discover corrections that enhance robustness rather than to reproduce the gate from scratch.

%\subsection{Observation and action spaces}

%The observation provided to the reinforcement-learning agent consists of a two-dimensional real-valued vector, which encodes the normalized static parameter offsets $(\delta\omega_1,\delta\omega_2)$ of the device. The bounds are chosen to safely cover the range of fluctuations considered in the noise model and ensure well-scaled inputs during training.

%The action produced by the agent is a $2K$-dimensional vector $\bm{a} = [-1,1]^{2K}$, whose components parameterize the cosine-basis residual corrections for the two control drives. These normalized actions are mapped linearly to the corresponding residual coefficients, providing a bounded and well-conditioned control interface.

%Together, these choices define a low-dimensional, continuous observation-action mapping in which the agent learns to associate device-specific imperfections with structured pulse corrections.

% ======================================================
%  DEEP REINFORCEMENT LEARNING AGENTS
% ======================================================
\section{Deep reinforcement learning algorithms}
\label{sec:DRL_Algo}

%\subsection{Algorithms and hyperparameters}

\begin{figure}[h]
    %\centering
    \includegraphics[width=\columnwidth]{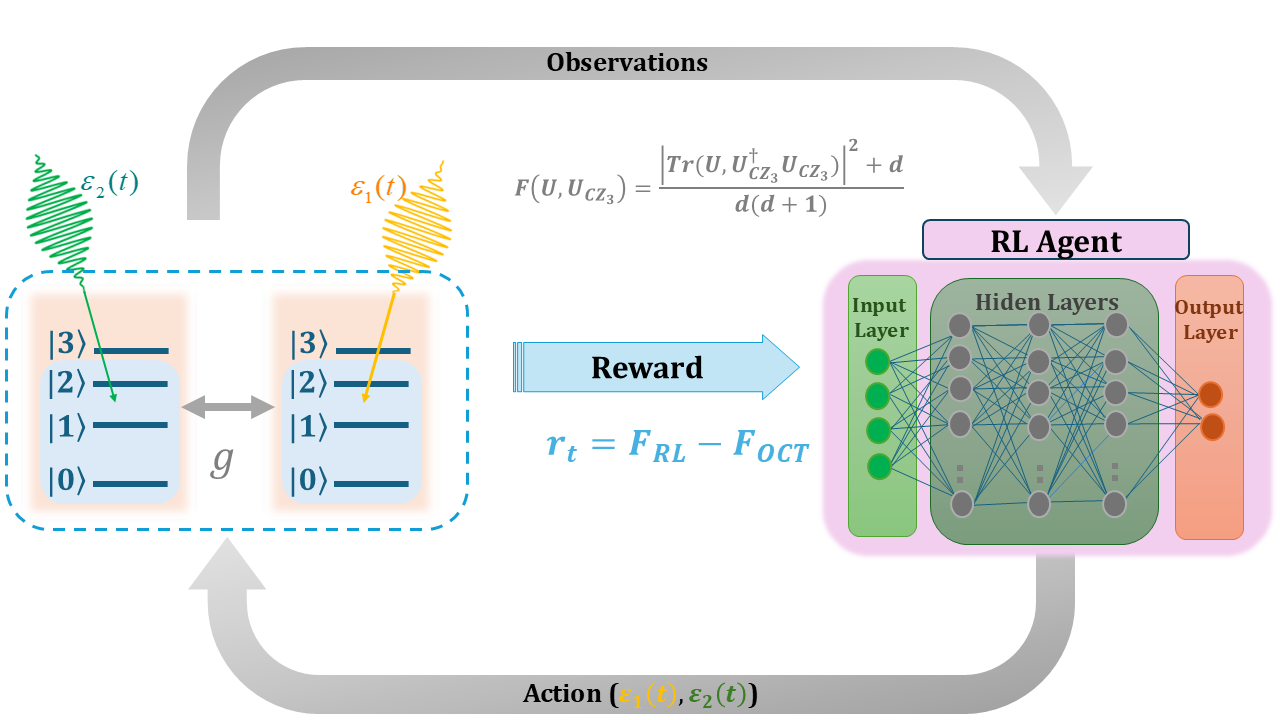}
    \caption{
    DRL workflow for residual pulse optimization.}
    \label{fig:DRL_workflow}
\end{figure}

Figure~\ref{fig:DRL_workflow} summarizes the DRL
workflow used in this work. At each training step, a noisy device
instance is generated by sampling static parameter offsets
$(\delta\omega_1,\delta\omega_2)$ from the prescribed Gaussian
distributions. These offsets constitute the contextual observation provided to the agent. 
Conditioned on this context, the agent outputs a
low-dimensional set of continuous coefficients that parametrize smooth residual corrections added to the QOCT baseline pulses, rather than redesigning the control pulse from scratch. The corrected
pulses are applied to the device Hamiltonian, the resulting gate
fidelity is evaluated, and a scalar reward defined as the incremental fidelity gain relative to the QOCT pulse on the same device—is returned to the agent. Since each episode consists of a single device evaluation,
the learning task naturally takes the form of a contextual bandit.

We compare four deep reinforcement-learning algorithms for continuous
control: Soft Actor--Critic (SAC)~\cite{Haarnoja2018}, Twin-Delayed Deep
Deterministic Policy Gradient (TD3)~\cite{Fujimoto2018}, Deep
Deterministic Policy Gradient (DDPG)~\cite{Lillicrap2015}, and Proximal
Policy Optimization (PPO)~\cite{Schulman2017}.
We consider these four algorithms as they constitute a representative and widely adopted set of DRL methods for continuous control. Together, they span complementary design choices, including on-policy versus off-policy learning (PPO versus SAC, TD3, and DDPG), deterministic versus stochastic policies, and different strategies for stabilizing training in continuous action spaces. In addition, to verify that the observed behavior does not depend on the specific choice of learning algorithm. Such comparative evaluations across multiple standard algorithms are common practice in the DRL literature, as no single method is known to be universally optimal across tasks. 

All agents employ identical multilayer perceptrons for the policy and
value networks, with two hidden layers of 256 units and ReLU activations.
A learning rate $\eta = 3\times 10^{-4}$ and discount factor
$\gamma = 0.99$ are used throughout.
These chosen hyperparameters follow standard settings widely used in continuous-control benchmarks. In particular, the learning rate $\eta$ and the discount factor $\gamma$
provide a good balance between training stability and convergence speed. The network architecture (two hidden layers of 256 units) is sufficient to represent the low-dimensional policy required for the cosine-basis parametrization while avoiding overparameterization. We verified that moderate variations of these hyperparameters do not qualitatively affect the results, indicating that the observed performance is consistent with respect to their precise values.

For off-policy methods (SAC, TD3,
DDPG), target networks are updated using a soft update with coefficient
$\tau = 0.005$. Exploration in SAC, TD3, and DDPG is implemented via
Gaussian action noise, while PPO relies on a stochastic policy with
clipped policy updates.

Each algorithm is trained for $T_{\mathrm{train}} = 10^5$ independent environment interactions,
with each interaction corresponding to a single sampled noisy device and a single pulse evaluation.
This training budget is sufficient to ensure convergence of the learning curves while remaining computationally tractable. Once training is completed offline, adapting the control pulse to a new
device requires only a single forward pass of the trained policy network,
making the computational cost of this approach negligible compared to
re-running QOCT.

%\subsection{Logging and evaluation}

%During training, we log the gate fidelity $F_{\mathrm{DRL}}$ obtained at the end of each episode. Since each episode corresponds to a single device evaluation, this yields a sequence of fidelities indexed by training iterations. From these data, we compute the per-episode fidelity and the
%running best fidelity.

After training, each agent is evaluated in three distinct settings:
\begin{enumerate}
    \item \emph{Nominal device:} zero parameter offsets,
    $(\delta\omega_1,\delta\omega_2)=(0,0)$.
    \item \emph{Single static-noise device:} a randomly selected triplet
    $(\delta\omega_1,\delta\omega_2)$ held fixed across all methods.
    \item \emph{Ensemble of noisy devices:} $M=100$ independently sampled devices drawn from the same noise distribution used during training.
\end{enumerate}

For each setting, we compute the average gate fidelity, and for the ensemble case we additionally report the standard deviation across devices. These metrics allow for a direct and quantitative comparison between the transfer performance of the nominal QOCT pulse and its improvement through
context-conditioned DRL corrections.

% ======================================================
% VI. NUMERICAL RESULTS
% ======================================================
\section{Numerical results}
\label{sec:results}

\subsection{Nominal Device}

\begin{figure}[h]
    %\centering
    \includegraphics[width=0.8\columnwidth]{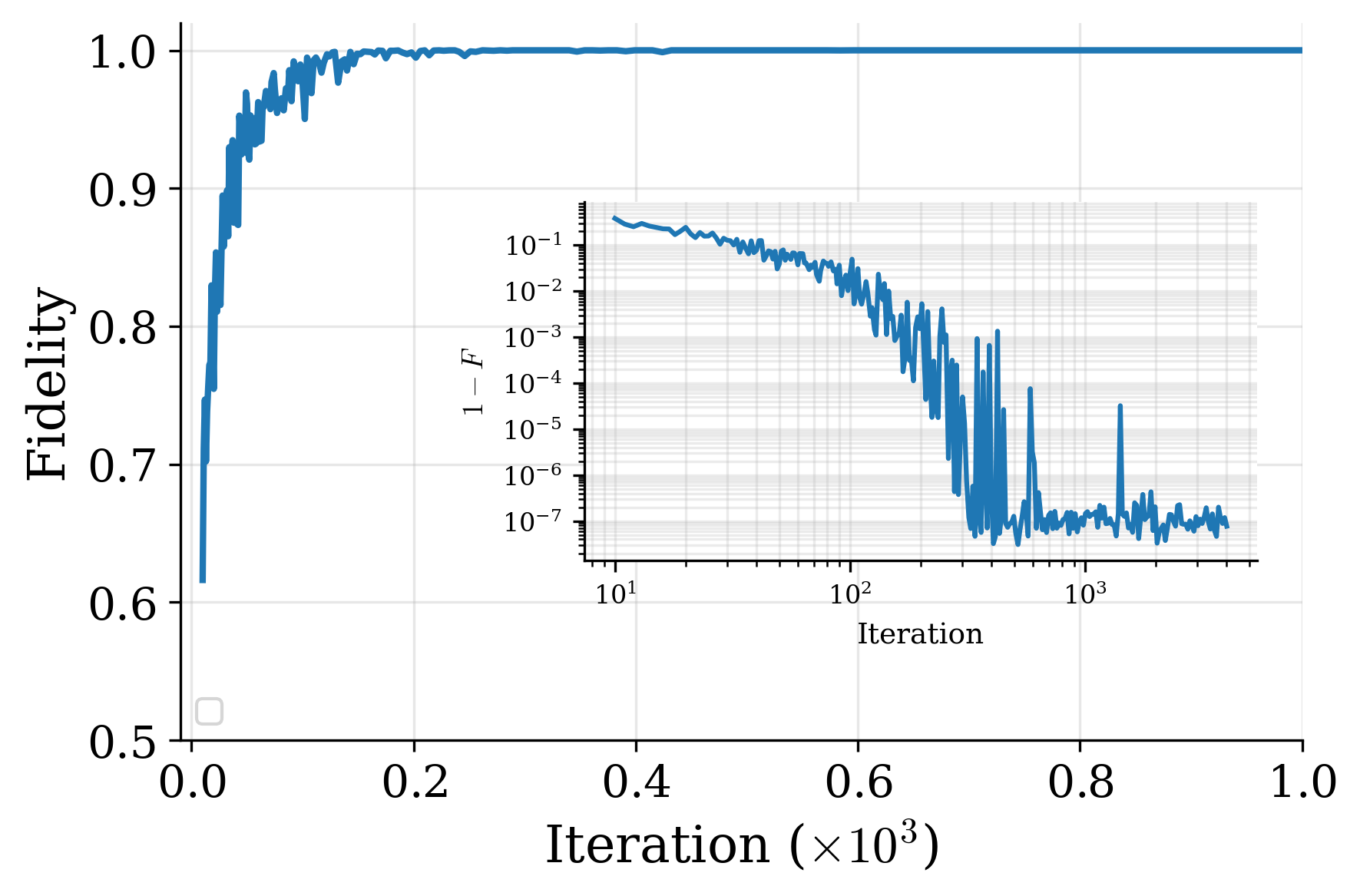}
    \caption{Convergence of the GRAPE optimization on the nominal two-qutrit Hamiltonian with a final infidelity of $8.1\times 10^{-8}$.}
    \label{fig:QOCT_iter}
\end{figure}

In the simulations reported here, the GRAPE optimization was performed
in the full two-qutrit Hilbert space of dimension $d=9$, using a total
gate duration $T = 1600$.
We find that for this duration, the optimizer reliably converges to
near-unit process fidelity, as shown in Fig.~\ref{fig:QOCT_iter}.
By contrast, for short total times ($T \lesssim 628$ in our units),
the optimization systematically fails to reach high fidelity, even
when increasing the number of iterations or varying the initial
guess pulses. This behavior is consistent with the presence of a
minimum gate time imposed by the coupled-transmon Hamiltonian. Such a minimum-time threshold is not an artifact of the numerical
optimization but reflects a fundamental constraint on two-qutrit
entangling gates. Indeed, in a recent study, the authors have 
analyzed quantum speed limits for two-qutrit controlled-phase gates
using both optimal-control constructions and analytical bounds,
showing that high-fidelity implementations are only possible above
a system-dependent minimum duration~\cite{Basyildiz2025}.
Our choice of $T=1600$ is therefore well above the minimum gate time set by speed-limit regime, ensuring that the nominal QOCT solution
saturates the accessible control landscape.

We will use these nominal pulses as a baseline against which both noisy-device performance and DRL-corrected pulses are compared.
\vspace{0.2cm}

For the nominal device, Figure~\ref{fig:QOCT_iter} shows the convergence of the process fidelity as a function of iteration number, illustrating the rapid ascent to near-unity fidelity within a few hundred iterations. From the perspective of control theory, this demonstrates that the GRAPE solution essentially saturates the reachable optimum on the nominal Hamiltonian.
On the nominal device, where the system Hamiltonian is perfectly known and free from noise or parameter uncertainty, QOCT provides an upper bound on the achievable gate fidelity. As shown in Fig.~\ref{fig:QOCT_iter}, the QOCT-optimized control pulse achieves a final process infidelity of  $(1-F_{\mathrm{nom}}^{\mathrm{QOCT}} )= 8.147\times 10^{-8}$
 and remains constant throughout the optimization horizon. This behavior reflects the deterministic nature of QOCT, which exploits full model knowledge and analytic gradients to converge directly toward a globally optimal solution.
Because the nominal device exactly matches the assumptions used in the QOCT optimization, there is no model mismatch, stochasticity, or partial observability. As a result, QOCT efficiently suppresses coherent control errors, including unwanted population transfer within the two-qutrit Hilbert space and residual phase errors, yielding a near-perfect implementation of the target CZ gate. The QOCT result therefore serves as a theoretical benchmark against which learning-based approaches can be quantitatively compared.

\begin{figure}[h]
    %\centering
    \includegraphics[width=0.8\columnwidth]{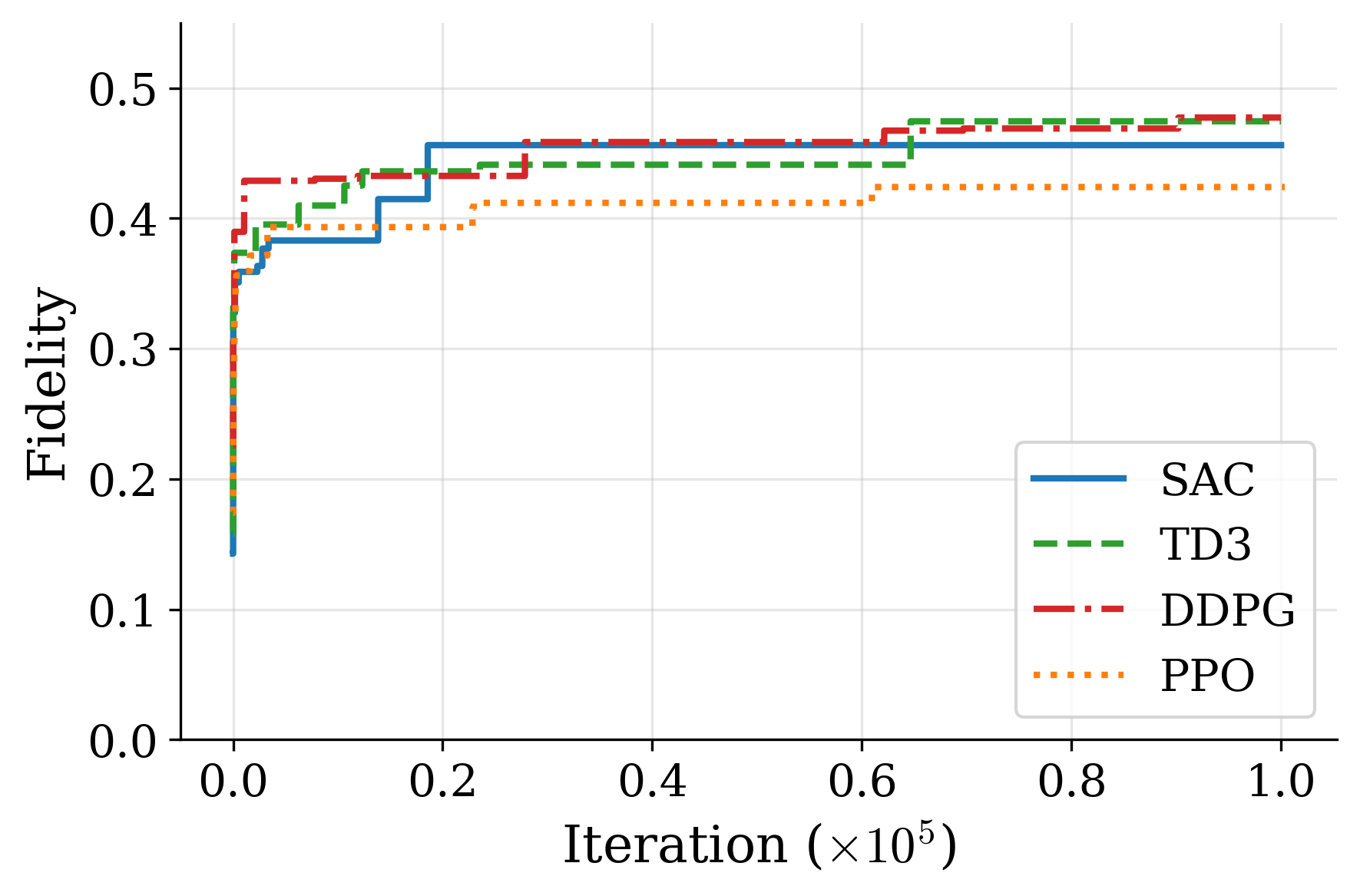}
   \caption{
    Learning curves on the nominal device comparing DRL algorithms (DDPG, TD3, SAC, PPO).}
    \label{fig:nominal_learning_curves}
\end{figure}

In contrast, if we apply DRL methods to perform the same pulse optimization task from scratch, these methods display qualitatively poorer performance. Figure~\ref{fig:nominal_learning_curves} shows the learning curves for DDPG, TD3, SAC, and PPO. All agents exhibit a rapid initial improvement in fidelity during early training, followed by a pronounced plateau at moderate fidelity values ($\approx0.4 - 0.48)$, well below the near-unit fidelity achieved by QOCT.

Despite extended training up to $10^5$ iterations, none of the DRL methods approach the QOCT benchmark. This performance gap is intrinsic to the DRL formulation: the agent must infer optimal control pulses solely from scalar reward signals derived from fidelity evaluations, without access to analytic gradients or exact dynamical information. Even on a nominal device, the high-dimensional continuous action space and the highly non-linear dependence of the final gate fidelity on hundreds of pulse-amplitude parameters significantly limit the achievable fidelity within feasible training times.
This challenge is further compounded by the sparse reward structure associated with high-fidelity gate synthesis, where meaningful rewards are obtained only when the generated evolution closely approximates the target two-qutrit entangling operation.

The plateau near a fidelity of approximately $\approx0.4 - 0.48$ reflects the fact that the DRL agents converge to pulse policies that produce structured but incorrect unitary operations within the computational subspace. In other words, the learned pulses generate dynamics that are nontrivial but do not realize the desired entangling transformation.
This behavior illustrates the difficulty of discovering high-fidelity solutions in the high-dimensional continuous pulse space using model-free learning alone. 
While DRL has been successfully applied to two-qubit gate design, including recent demonstrations on superconducting transmon systems~\cite{Nguyen2024DRLTransmon}, these results are obtained in carefully designed control settings. Applying fully model-free DRL directly to high-dimensional, time-discretized control fields remains challenging. This difficulty is further observed in the present setting, which involves a two-qutrit system rather than a two-qubit system. Furthermore, although the increase in system size from two qubits to two qutrits may appear modest, the Hilbert-space dimension grows from four to nine and the number of dynamical constraints increases accordingly, making the control landscape significantly more challenging to explore for model-free DRL.

\subsection{QOCT-enhanced DRL on the nominal device}

\begin{figure}[h]
    %\centering
    \includegraphics[width=0.8\columnwidth]{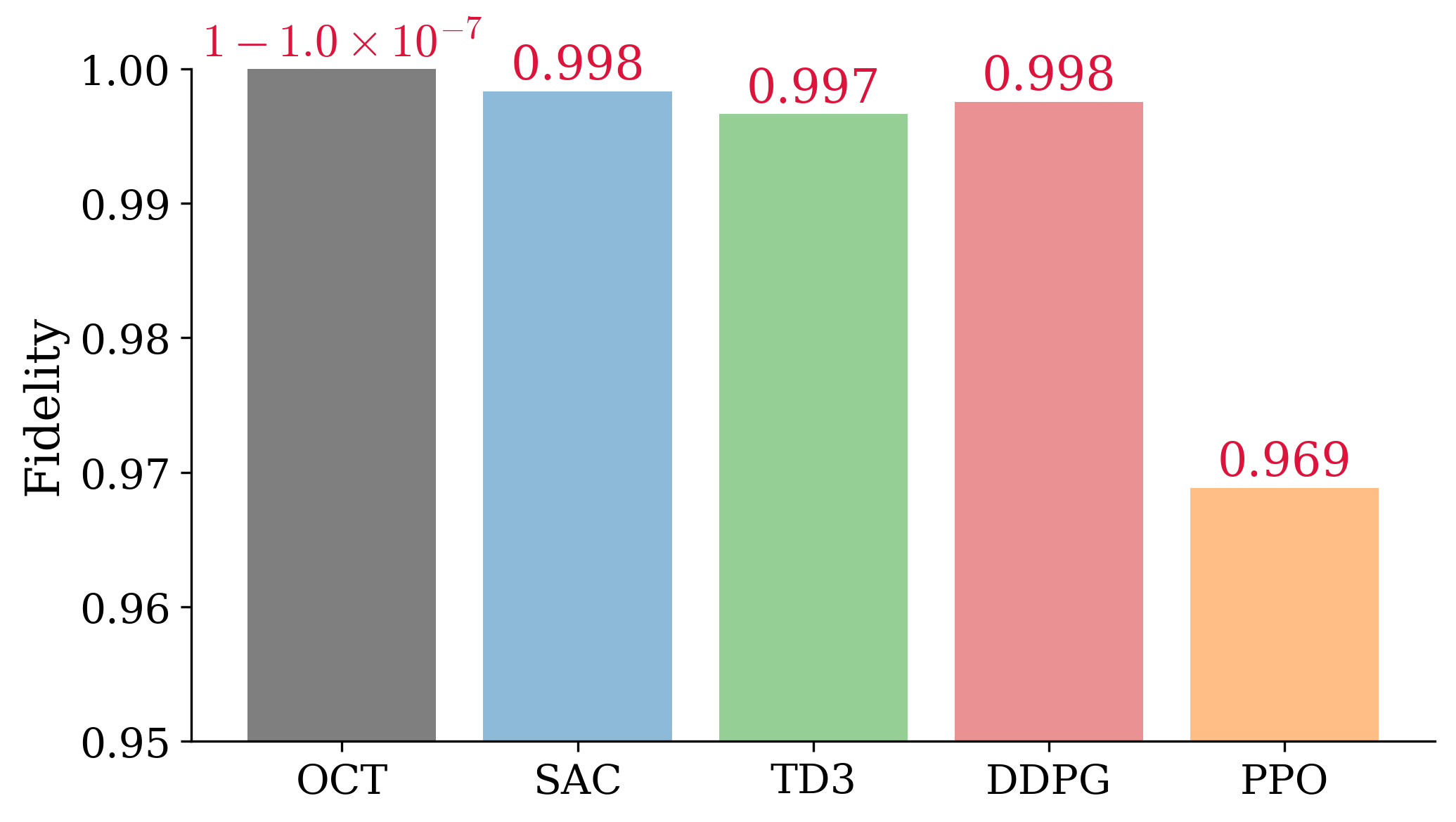}
    \caption{Average gate fidelity on the nominal device for QOCT and DRL-enhanced pulses.}% QOCT reaches $F = 1 - 1.0\times 10^{-7}$, while the DRL-corrected pulses for SAC, TD3 and DDPG preserve near-unit fidelity ($>0.994$), and PPO reaches $0.970$.}
    \label{fig:nominal_bar}
\end{figure}

While standalone DRL fails to reach high fidelities on the nominal device, its performance changes qualitatively once QOCT-optimized pulses are provided as an initialization. Figure~\ref{fig:nominal_bar} reports the gate nominal fidelities obtained after DRL refinement, together with the QOCT reference. All QOCT-initialized DRL agents preserve high performance on the nominal device, with SAC and DDPG remaining extremely close to the QOCT solution (0.998), TD3 exhibiting a small degradation (0.997), and PPO showing the largest deviation (0.969).
Importantly, none of the DRL-enhanced pulses surpass the QOCT fidelity on the nominal Hamiltonian. This observation is consistent with the saturation of the quantum control landscape when the system model is exact: once an optimal solution is reached by QOCT, further learning cannot improve performance in the absence of model mismatch. Instead, DRL introduces small perturbations around the QOCT optimum, which may slightly reduce the nominal fidelity.

These results clarify the respective roles of QOCT and DRL. On the nominal device, QOCT remains optimal and defines a strict upper bound on achievable fidelity. DRL does not improve performance in this idealized setting; rather, its role is to preserve near-optimal control while enabling robustness and adaptability under parameter uncertainty. Maintaining near-unit fidelity on the nominal model is therefore not the objective of the DRL layer, since QOCT already achieves this limit. Rather, it serves as a baseline requirement that ensures the learned corrections preserve performance when applied to devices with static parameter mismatches.

\subsection{Single static-noise device}

We now consider a single frequency-disordered device generated by introducing
independent Gaussian fluctuations in the two qutrit frequencies. Thus, the applied frequency fluctuations correspond to relative errors of
approximately $10^{-3}$ for the first qutrit and $1.1\times10^{-3}$ for the
second qutrit.
When applied directly to this noisy Hamiltonian, the baseline QOCT pulse-optimized for the nominal device exhibits a substantial degradation in performance, reaching a fidelity of approximately $0.92$. This reduced fidelity reflects the transfer performance of a nominally optimized
pulse under static model mismatch, rather than a failure of QOCT to optimize the Hamiltonian when it is accurately known.

\begin{figure}[h]
    %\centering
    \includegraphics[width=0.8\columnwidth]{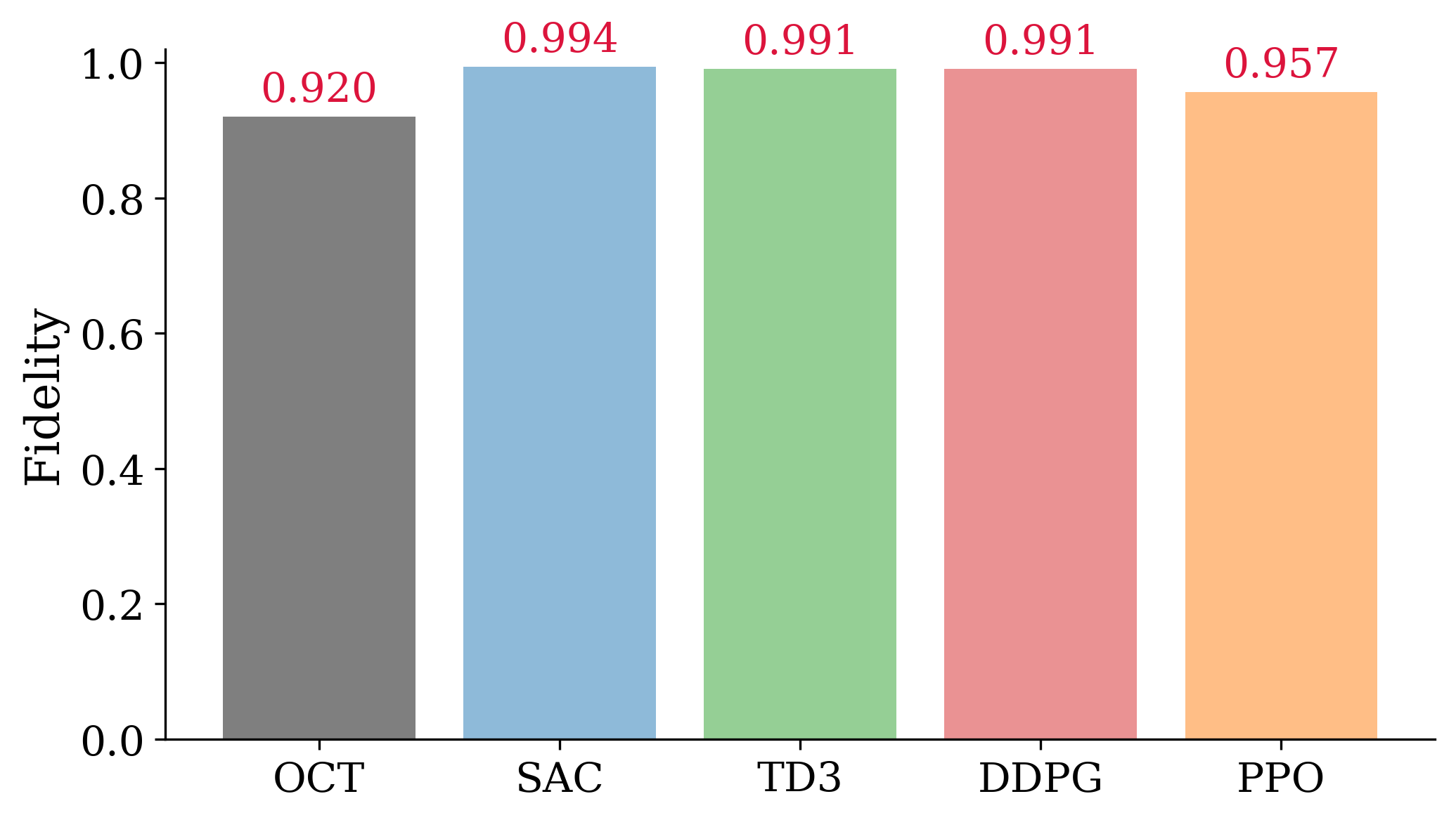}
    \caption{ Gate fidelity on a single frequency-disordered device with Gaussian frequency fluctuations of  $\sigma_\omega=10^{-3}$.}
    \label{fig:single_bar}
\end{figure}

Fig.~\ref{fig:single_bar} compares the fidelity obtained by the nominal QOCT pulse and by the
DRL-enhanced pulses on the same frequency-disordered device. The DRL-enhanced pulses substantially recover the lost fidelity. In fact, SAC agent reaches $F_{\rm SAC}=0.994$, while TD3 and DDPG both reach $F_{\rm TD3}=F_{\rm DDPG}=0.991$.
PPO also improves over the QOCT baseline, reaching $F_{\rm PPO}=0.957$. These results show that small context-dependent residual corrections learned
by DRL can efficiently compensate static frequency detuning errors without
requiring a full re-optimization of the QOCT pulse.

This behavior stands in sharp contrast to the nominal-device case, where DRL was unable to match or surpass the QOCT benchmark. Here, the presence of static model mismatch highlights the sensitivity of nominal
open-loop control to parameter inaccuracies and the potential of calibration-based
approaches to mitigate this sensitivity.
As a result, the learned residual corrections enable recovery of high gate fidelity
without re-optimizing the full control pulse for the specific device realization. Furthermore, once trained, the DRL policy can generate correction pulses through a single forward pass of the neural network, avoiding repeated high-dimensional optimization for each device realization.
 
Among the algorithms considered, DDPG and TD3 efficiently exploit the smooth, low-dimensional residual parameterization to learn deterministic pulse corrections, with TD3 exhibiting slightly more stable convergence due to its reduced critic bias. SAC achieves similarly high fidelities while benefiting from entropy-regularized exploration, which enhances robustness to model mismatch. PPO, although more stable due to its on-policy updates, converges to slightly lower fidelities in this continuous control setting.

\subsection{Ensemble robustness and noise statistics}
\label{subsec:ensemble}

We next assess robustness by evaluating QOCT-only and DRL-enhanced pulses on an ensemble of \(M=100\) noisy devices drawn from the same Gaussian parameter distributions used during training. For each method, we compute the ensemble-averaged fidelity \(\overline{F}_{\mathrm{ens}}\) together with its standard deviation, which quantifies performance variability across devices.

\begin{figure}[h]
    %\centering
    \includegraphics[width=0.8\columnwidth]{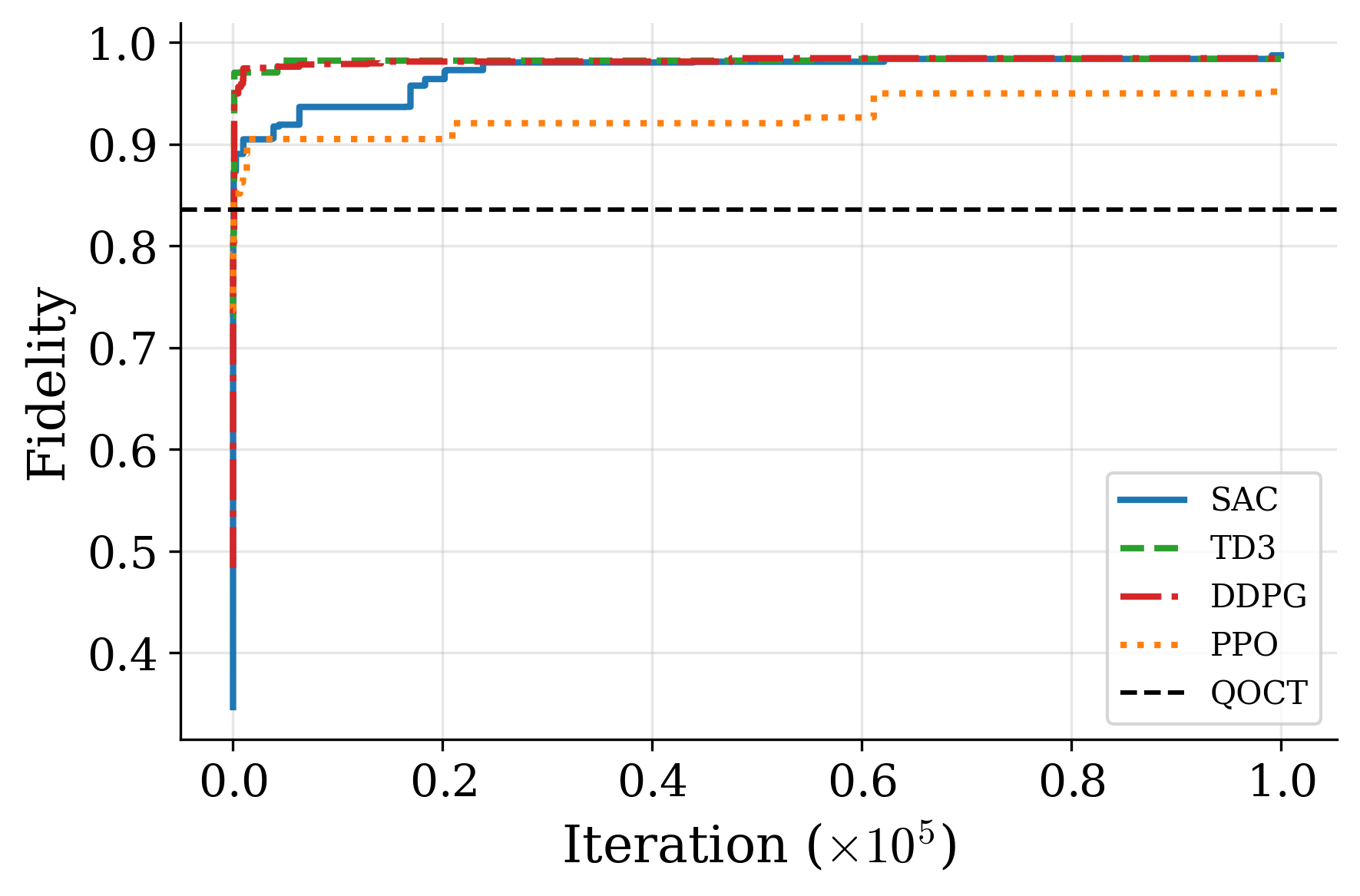}
    \caption{Training curves for SAC, TD3, DDPG, and PPO in the contextual bandit environment. For each algorithm we plot the gate fidelity. The horizontal solid blue line indicates the QOCT ensemble mean fidelity $F^{\mathrm{QOCT}}_{\mathrm{ens}} = 0.836$.}
    \label{fig:learning}
\end{figure}

Figure~\ref{fig:learning} shows the training curves for SAC, TD3, DDPG, and PPO in the contextual bandit setting. All algorithms exhibit a rapid initial increase in fidelity, reflecting their ability to exploit the nominal QOCT pulse as a strong starting point. Subsequent improvements occur more gradually and correspond to fine pulse adjustments that enhance robustness across the ensemble of static Hamiltonian perturbations. TD3 and DDPG converge most rapidly, reaching near-unit fidelities within \(\mathcal{O}(10^4)\) episodes, while SAC converges more gradually but remains highly stable throughout training. PPO improves steadily but saturates at a lower final fidelity. The horizontal reference line indicates the QOCT ensemble-averaged fidelity, $F_{\rm ens}^{\rm QOCT}=0.836$, which remains significantly below the fidelities reached by the DRL-enhanced
methods after training

\begin{figure}[h]
    %\centering
    \includegraphics[width=0.8\columnwidth]{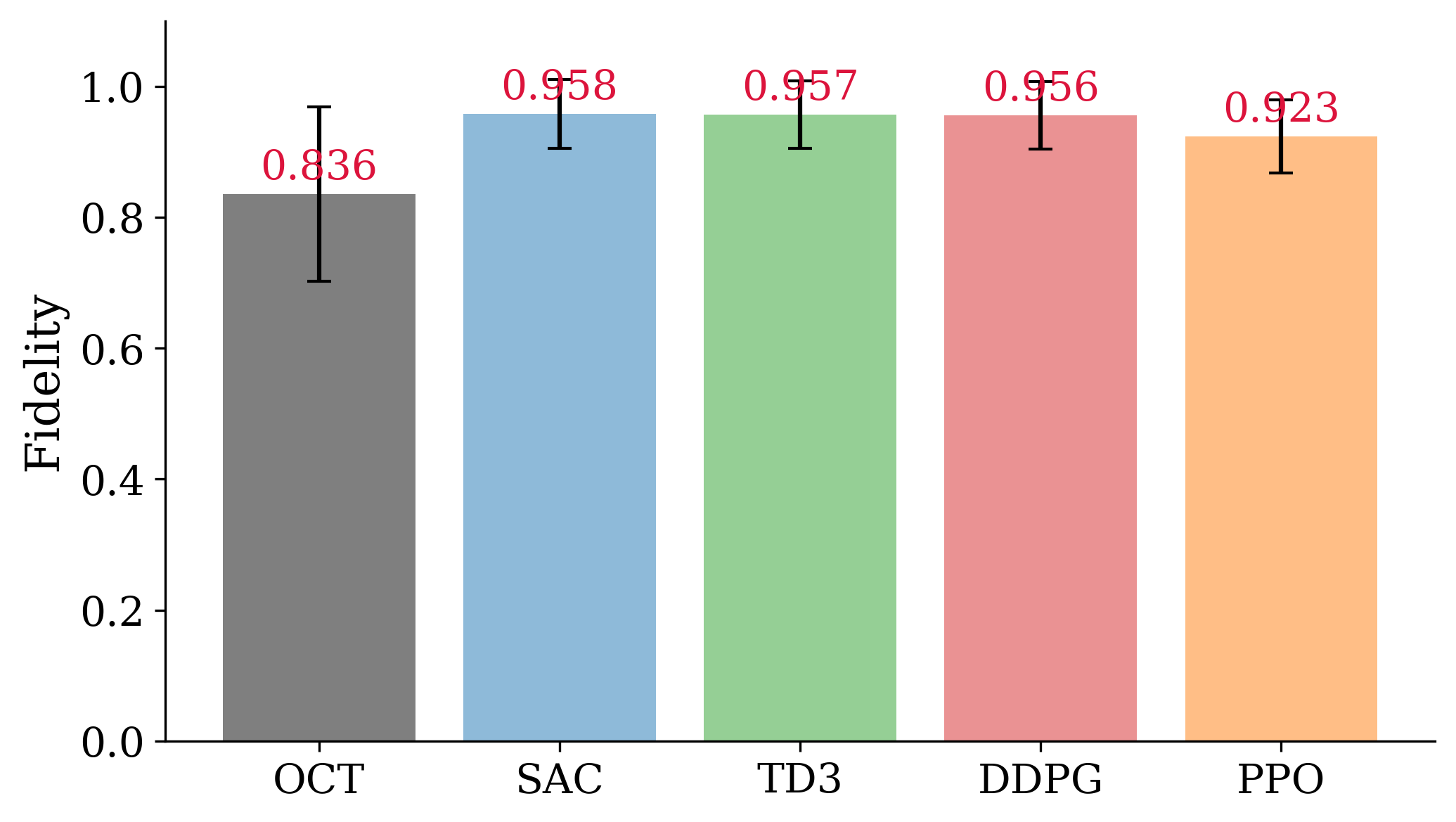}
    \caption{Ensemble-averaged fidelity over 100 frequency-disordered devices with $\sigma_\omega=10^{-3}$. Error bars indicate the standard deviations across the ensemble.}
    %The nominal QOCT pulse reaches an average fidelity of $0.836$, while the DRL-enhanced pulses improve the ensemble performance to $0.958$ for SAC, $0.957$ for TD3, $0.956$ for DDPG, and $0.923$ for PPO. Error bars indicate one standard deviation across the ensemble.}}
    \label{fig:ensemble_bar}
\end{figure}

The ensemble statistics are summarized in Fig.\ref{fig:ensemble_bar}. The nominal QOCT pulse
achieves an ensemble-averaged fidelity
$F_{\rm ens}^{\rm QOCT}=0.836$.
The DRL-enhanced pulses substantially improve the ensemble performance, with $F_{\rm ens}^{\rm SAC}=0.958$,
$F_{\rm ens}^{\rm TD3}=0.957$, 
$F_{\rm ens}^{\rm DDPG}=0.956$, and $F_{\rm ens}^{\rm PPO}=0.923$.
Thus, all DRL-enhanced methods outperform the nominal QOCT pulse across the
frequency-disordered ensemble, with SAC giving the best average fidelity.

As in the single-device case, the advantage of DRL over
the nominal QOCT baseline arises from its ability to learn device-dependent
residual corrections that mitigate the sensitivity of open-loop control to
static model mismatch.
Across the ensemble, each realization corresponds to a distinct static Hamiltonian for which the nominal QOCT pulse is generally suboptimal. Here, the QOCT ensemble performance reflects the transfer fidelity of a
single pulse optimized on the nominal Hamiltonian and then applied without
re-optimization to many distinct static Hamiltonian realizations.
By contrast, the DRL agents learn context-conditioned residual corrections
that improve the transfer performance of the same nominal pulse across the
distribution of parameter offsets.
This leads simultaneously to higher average fidelity and a substantial reduction in performance variability across devices.

\vspace{0.5cm}
We emphasize that this ensemble analysis should be distinguished from ensemble-robust optimal control, in which the control pulse is obtained by explicitly maximizing the average fidelity over a known parameter distribution, as demonstrated for example by Poggi \emph{et al.}~\cite{Poggi2024}. 
Ensemble-robust QOCT represents a powerful open-loop robustness strategy, but it assumes prior knowledge of the full disorder distribution and requires substantially higher computational cost. The present work instead focuses on a complementary regime, where a single nominal pulse is deployed and lightweight, context-dependent corrections are learned to adapt to individual devices without repeated re-optimization.

%Relative to the nominal QOCT baseline, which exhibits both a low mean transfer fidelity and a large spread all DRL agents substantially improve robustness. SAC achieves the highest average fidelity with the smallest variance, while TD3 and DDPG provide comparable robustness with faster convergence. PPO also significantly outperforms QOCT but remains less robust than the off-policy methods. Overall, the standard deviation of the fidelity is reduced from \(0.128\) for QOCT to approximately \(0.012\) for the best-performing DRL agents, indicating a much more uniform and reliable gate performance across the ensemble. This reduction in variance highlights the suitability of the hybrid QOCT+DRL approach as a scalable calibration strategy for multi-device platforms.

\subsection{Robustness to imperfect parameter estimation.}

In realistic experimental settings, the device parameters entering the control model are not known exactly. They are inferred from calibration measurements and therefore contain statistical uncertainty, with the achievable estimation accuracy limited by measurement noise and finite coherence times ($T_1, T_2$), which constrain the precision of parameter estimation. To assess the impact of such imperfections, we model the contextual inputs provided to the DRL agent as noisy estimates of the underlying parameter deviations. The agent observes a vector of normalized offsets as:
\[
\tilde{o} = \left(
\frac{\delta\omega_1 + \epsilon_{\omega_1}}{\sigma_\omega},
\frac{\delta\omega_2 + \epsilon_{\omega_2}}{\sigma_\omega}
%\frac{\delta g + \epsilon_g}{\sigma_g}
\right),
\]

where the estimation errors are modeled as independent Gaussian variables, $\epsilon_{\omega_1}, \epsilon_{\omega_2} \sim \mathcal{N}(0,\eta_\omega^2)$. The quantities \(\delta\omega_1\) and \(\delta\omega_2\) represent the true physical mismatch between the device and the nominal model, with characteristic scales set by \(\sigma_\omega\), as studied previously. In contrast, the quantities \(\epsilon_{\omega_1}\) and \(\epsilon_{\omega_2}\) represent errors in the estimation of these parameter deviations, arising from the finite accuracy of calibration procedures. These estimation errors affect only the information available to the DRL agent, while the system dynamics are always governed by the true parameter deviations.

The estimation error is characterized by the dimensionless ratios \(\eta_\omega/\sigma_\omega\), where \(\sigma_\omega\) denotes the characteristic width of the underlying device-parameter fluctuations.. We consider values ranging from ($10^{-4}$ to $1$, spanning regimes from near-perfect parameter estimation to observation errors comparable to the underlying device variability itself.

\begin{figure}[h]
%\centering
\includegraphics[width=\columnwidth]{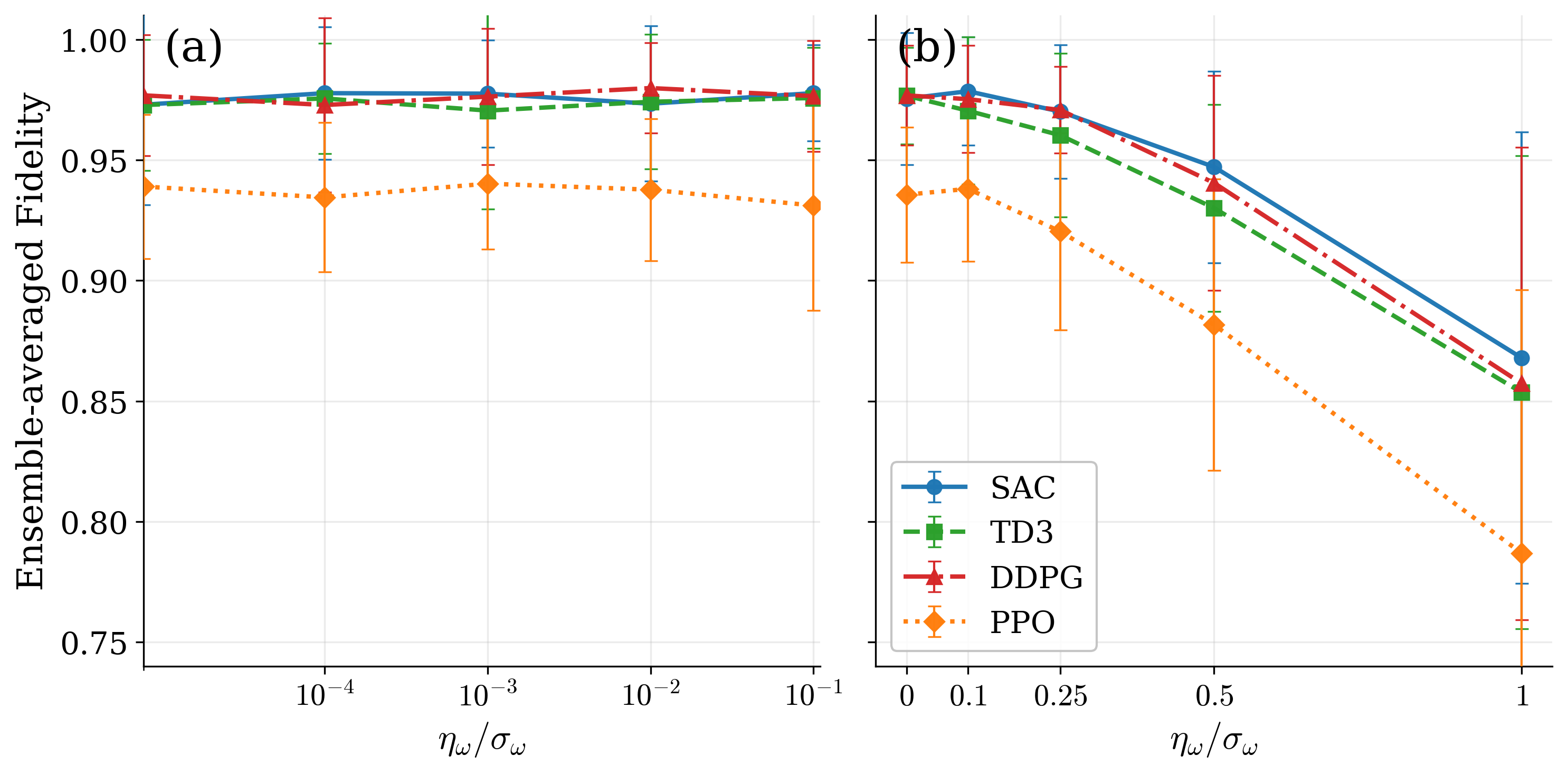}
\caption{
Robustness of the QOCT-DRL calibration protocol to imperfect frequency estimation. The DRL agents are evaluated on the same ensemble of frequency-disordered devices, while the contextual inputs are corrupted by Gaussian observation noise of strength \(\eta_\omega\). Panel (a) explores the regime of small estimation errors \((10^{-4}\le \eta_\omega/\sigma_\omega \le 10^{-1})\), whereas panel (b) extends the analysis to observation-noise levels comparable to the intrinsic device variability.
}
\label{fig:freq_obs_noise}
\end{figure}

The resulting performance is shown in Fig.~\ref{fig:freq_obs_noise}. Panel~(a) examines the regime of small estimation errors, spanning four orders of magnitude from \(\eta_\omega/\sigma_\omega=10^{-4}\) to ($10^{-1}$). Over this entire range, the ensemble-averaged fidelities obtained with SAC, TD3, and DDPG remain nearly unchanged and stay close to the ideal-observation value. Even PPO, which consistently yields lower fidelities, exhibits only a weak dependence on the estimation uncertainty. These results indicate that the learned calibration policies do not require highly accurate knowledge of the frequency deviations and can operate effectively using only coarse parameter estimates.

Panel~(b) extends the analysis to larger observation-noise levels. A gradual degradation of performance becomes visible once the estimation uncertainty reaches a significant fraction of the intrinsic device variability. Nevertheless, SAC, TD3, and DDPG continue to outperform the nominal QOCT pulse even when the observation noise becomes comparable to the underlying frequency fluctuations \((\eta_\omega/\sigma_\omega \sim 1)\). In contrast, PPO exhibits a more pronounced sensitivity to imperfect parameter estimation and falls below the QOCT baseline in the largest-noise regime.

Overall, the results demonstrate that the proposed QOCT-DRL calibration framework is remarkably tolerant to imperfect frequency characterization. In particular, realistic calibration errors corresponding to \(\eta_\omega/\sigma_\omega \lesssim 0.1\) have only a minor impact on the achieved fidelities. Significant degradation appears only when the uncertainty in the estimated frequency offsets becomes comparable to the actual spread of frequency variations across the device ensemble. This behavior suggests that the learned correction policies rely primarily on coarse information about the local parameter regime rather than on highly precise parameter estimates.

% ======================================================
% VII. DISCUSSION
% ======================================================
\section{Discussion}
\label{sec:discussion}

\subsection{Hybrid open/closed-loop control interpretation}

The limitations of direct model-free DRL observed in our study are consistent with previous studies showing that exploration becomes increasingly difficult in high-dimensional quantum-control problems, particularly when the controls are represented by large numbers of time-discretized parameters \cite{Sivak2022_PRX,Baum2021_PRXQuantum}. Experimental implementations on superconducting processors have likewise demonstrated that DRL can successfully optimize control pulses, but often requires substantial training effort and carefully designed learning strategies to achieve high-fidelity solutions \cite{Baum2021_PRXQuantum}. Related works have further shown that model-free DRL benefits significantly from prior information, demonstrations, or near-optimal initializations that guide exploration and stabilize training \cite{Nguyen2024DRLTransmon,Li2025_DRLfD}.

Motivated by these observations, the hybrid QOCT-DRL approach adopted here first generates a high-fidelity nominal pulse using QOCT and subsequently employs DRL only to learn residual calibration corrections around this reference solution. The residual controls are represented using a truncated cosine basis, which substantially reduces the dimensionality of the optimization problem. For the present two-qutrit gate, direct pulse optimization would require treating the amplitudes of two control channels over 160 time segments as independent variables, corresponding to 320 optimization parameters. In contrast, the residual correction is parameterized by only $K=20$ cosine coefficients per control channel, resulting in a 40-dimensional action space. Consequently, the DRL agent no longer explores the full pulse landscape but instead searches for a compact correction around an already high-quality solution, leading to significantly more stable and efficient learning.

Our framework can be interpreted as a hybrid open/closed-loop control scheme. The initial QOCT step is purely model-based: it uses a nominal Hamiltonian and gradient-based optimization to find high-fidelity pulses. This corresponds to \emph{open-loop optimal control} in the standard sense~\cite{Glaser2015,Koch2022}. The DRL stage, by contrast, incorporates \emph{closed-loop} information about device-specific parameter variations: the context vector encodes deviations from the nominal model, and the reward is based on actual performance on that device.

Conceptually, this is closely related to adaptive hybrid optimal control for imprecisely characterized systems, where initial pulses are obtained from model-based QOCT and subsequently refined using experimental feedback~\cite{Egger2014}. Our contribution is to implement this hybrid idea with a contextual DRL agent that learns a generalizable mapping from device parameters to residual corrections, rather than performing a separate local re-optimization for each device instance. This generalization is particularly valuable when calibrating a family of devices drawn from a similar fabrication process. In contrast to per-device re-optimization, this approach amortizes the
calibration effort across devices, enabling rapid adaptation without repeated
high-dimensional optimal control runs. This objective differs from closed-loop optimization methods that iteratively refine pulse parameters for a single device realization. Here, the goal is instead to learn a calibration policy that can be applied directly to previously unseen devices without repeating the optimization procedure.

\subsection{Implications for superconducting qutrit hardware}

The numerical results presented above indicate that a relatively shallow DRL layer, operating in a low-dimensional cosine basis, is sufficient to substantially enhance the robustness of a two-qutrit ($\CZthree$) gate against static parameter mismatch. In particular, the DRL-enhanced pulses consistently achieve significantly higher fidelities than the nominal QOCT pulse across the ensemble of frequency-disordered devices considered in this work, while exhibiting a much smaller variability from one device realization to another.

These findings are directly relevant to recent experimental progress in superconducting-qutrit platforms. In particular, Goss $\emph{et al.}$ demonstrated high-fidelity two-qutrit entangling gates in fixed-frequency transmon devices, reporting process fidelities exceeding ($95\%$) for microwave-activated ($\mathrm{CZ}$) and ($\mathrm{CZ}^{\dagger}$) gates \cite{Goss2022}. More generally, recent advances in multi-level superconducting processors continue to increase the importance of calibration techniques capable of compensating for device-dependent parameter variations and imperfect Hamiltonian models \cite{Goss2022,Roy2023}. In this context, the proposed QOCT-DRL framework provides a promising strategy in which QOCT generates a high-fidelity nominal pulse and DRL learns compact residual corrections that improve robustness across a family of related devices. Extending this approach to open-system dynamics and experimental closed-loop calibration constitutes a natural direction for future work \cite{Roy2023}.

\subsection{Extensions and outlook}

Several extensions of the present framework are worth exploring. First, incorporating open-system dynamics through a Lindblad master equation would allow the calibration policy to account explicitly for decoherence and dissipation. Second, the approach could be extended to larger multi-qudit architectures, where the complexity of direct pulse optimization grows rapidly and calibration becomes increasingly challenging \cite{Ashhab2026}. Finally, integrating the proposed QOCT-DRL framework with experimental calibration loops on superconducting platforms would provide an important test of its practical utility \cite{Roy2023}.

Overall, the combination of physics-based QOCT and data-driven DRL provides a promising route toward robust and scalable control of multi-level quantum hardware.

% ======================================================
% VIII. CONCLUSION
% ======================================================
\section{Conclusion}
\label{sec:conclusion}

We have proposed and numerically validated a hybrid control framework for robust two-qutrit $\CZthree$ gates with a special emphasis on weakly anharmonic superconducting circuits. Starting from GRAPE-optimized pulses on a nominal device, we train contextual DRL agents to output small cosine-basis residual corrections conditioned on device parameter offsets. The reward is defined as the fidelity improvement over the QOCT baseline for each noisy device. Our simulations using SAC, TD3, DDPG, and PPO show that the best agents significantly enhance ensemble-averaged fidelity relative to the
nominal QOCT baseline while preserving high nominal performance. On the nominal Hamiltonian, none of the agents surpass the QOCT solution,
emphasizing that their benefit lies not in redesigning optimal pulses, but in
mitigating static model mismatch, device-to-device variability, and slow parameter
drifts.
From a conceptual standpoint, this constitutes a DRL-based realization of hybrid open/closed-loop optimal control for imprecisely characterized systems. This approach is complementary to ensemble-robust optimal control strategies,
which explicitly optimize pulses over parameter distributions but require prior
knowledge of those distributions and substantially higher computational costs.
 Practically, it points toward automated, device-aware calibration strategies for multi-level superconducting processors. In future work, we anticipate that similar hybrid QOCT+DRL schemes will play an
important role in bridging the gap between idealized pulse design and the
realities of noisy, heterogeneous quantum hardware by enabling fast,
calibration-efficient adaptation across devices.
\begin{acknowledgments}
We would like to thank Bora Basyildiz, Tomoko Fuse, Makoto Negoro and Shotaro Shirai for useful discussions. S.A. was supported by Japan’s Ministry of Education, Culture, Sports, Science and Technology (MEXT) Quantum Leap Flagship Program Grant No. JPMXS0120319794.
\end{acknowledgments}
% ======================================================
% Appendix A: Alternative two-qutrit controlled-phase gate
% ======================================================
\appendix

\section{Optimized pulses for $\CZthree$ gate}
\label{app:CZ3 pulses}

\begin{figure}[h]
   % \centering
    \includegraphics[width=0.8\columnwidth]{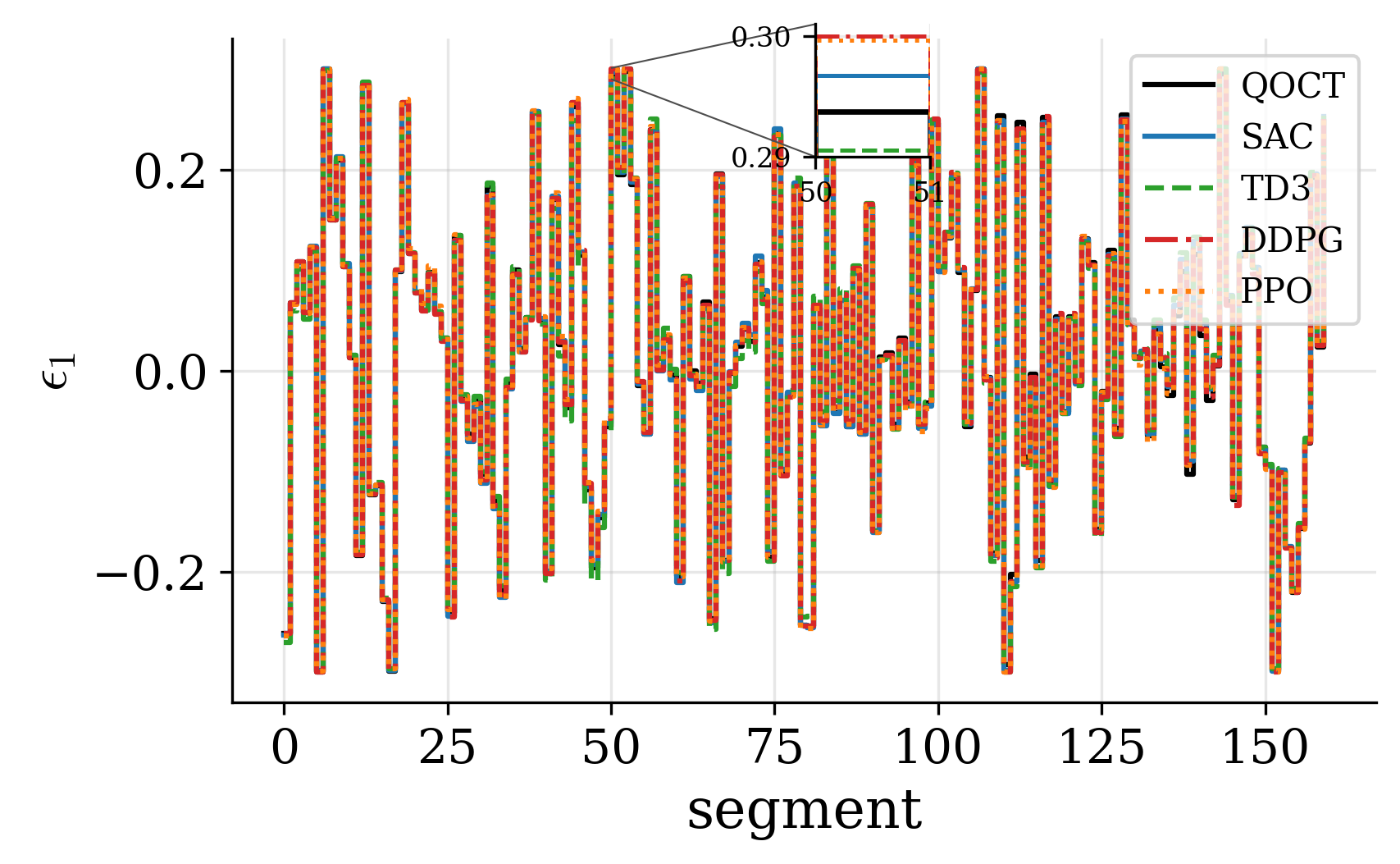}
    \caption{Drive \(\epsilon_1(t)\) on the static-noise device for QOCT (black) and for DRL-enhanced pulses obtained from DRL algorithms. The inset highlights the small but systematic deviations between the DRL-corrected pulses and the QOCT reference become visible.}
    \label{fig:pulses}
\end{figure}

%While the ensemble metrics quantify the robustness gains achieved by DRL, it is instructive to directly inspect the optimized control pulses. 
Figure~\ref{fig:pulses} shows the drive \(\epsilon_1(t)\) on a representative static-noise device for the QOCT baseline and the DRL-enhanced solutions.

Across the full gate duration, the DRL-optimized pulses remain extremely close to the QOCT reference and largely overlap at the scale of the main plot. The inset reveals that the corrections introduced by SAC, TD3, DDPG, and PPO consist of small, structured amplitude shifts relative to the QOCT pulse, typically at the few-percent level of the maximum drive amplitude. This underscores a key point: the substantial gains in fidelity and robustness observed in the noisy-device and ensemble settings arise from minimal yet well-placed modifications of the QOCT solution. In this sense, DRL acts as a fine calibration layer that compensates static parameter mismatches without introducing large-amplitude or highly oscillatory control features.
A similar behavior is observed for the second control channel \(\epsilon_2(t)\) on the same static-noise device. As for \(\epsilon_1(t)\), all DRL-enhanced pulses closely track the QOCT reference throughout the gate duration, with only small residual corrections visible upon close inspection.

\section{Alternative two-qutrit controlled-phase gate}
\label{app:alt_gate}
\begin{figure}[h]
    %\centering
    %\includegraphics[width=0.8\columnwidth]{OCT_fid_vs_iter_2.png}\\[0.5ex]
    \includegraphics[width=0.8\columnwidth]{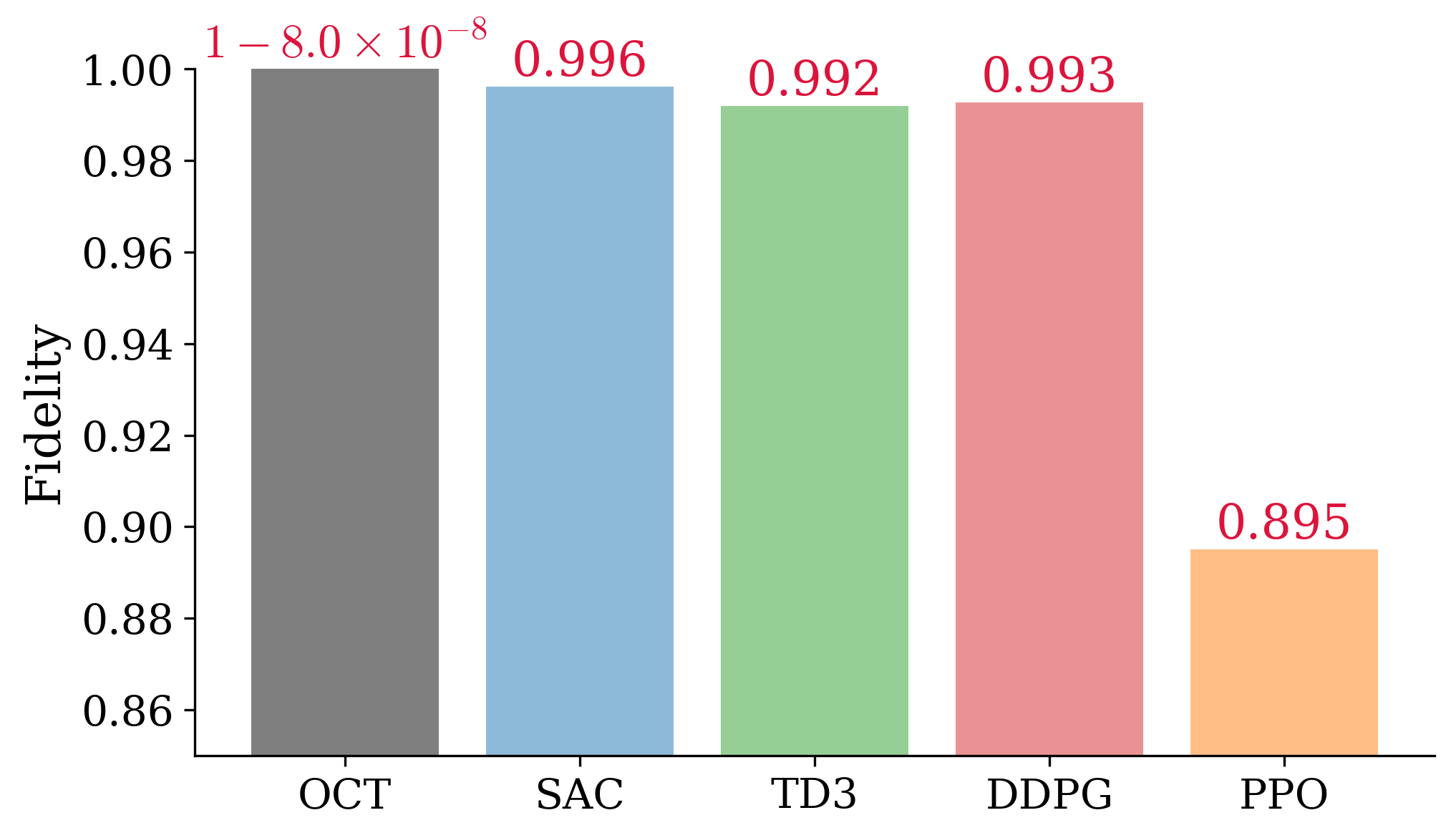}
    \caption{%Top: QOCT (GRAPE) convergence for $U_{\mathrm{alt}}$ on the nominal Hamiltonian. Bottom:
    Nominal-device fidelities for QOCT and DRL-enhanced pulses.}
    \label{fig:alt_QOCT_and_nominal}
\end{figure}

\begin{figure}[h]
  %  \centering
%    \includegraphics[width=\columnwidth]{noise_distribution_2.png}\\[0.5ex]
    \includegraphics[width=0.8\columnwidth]{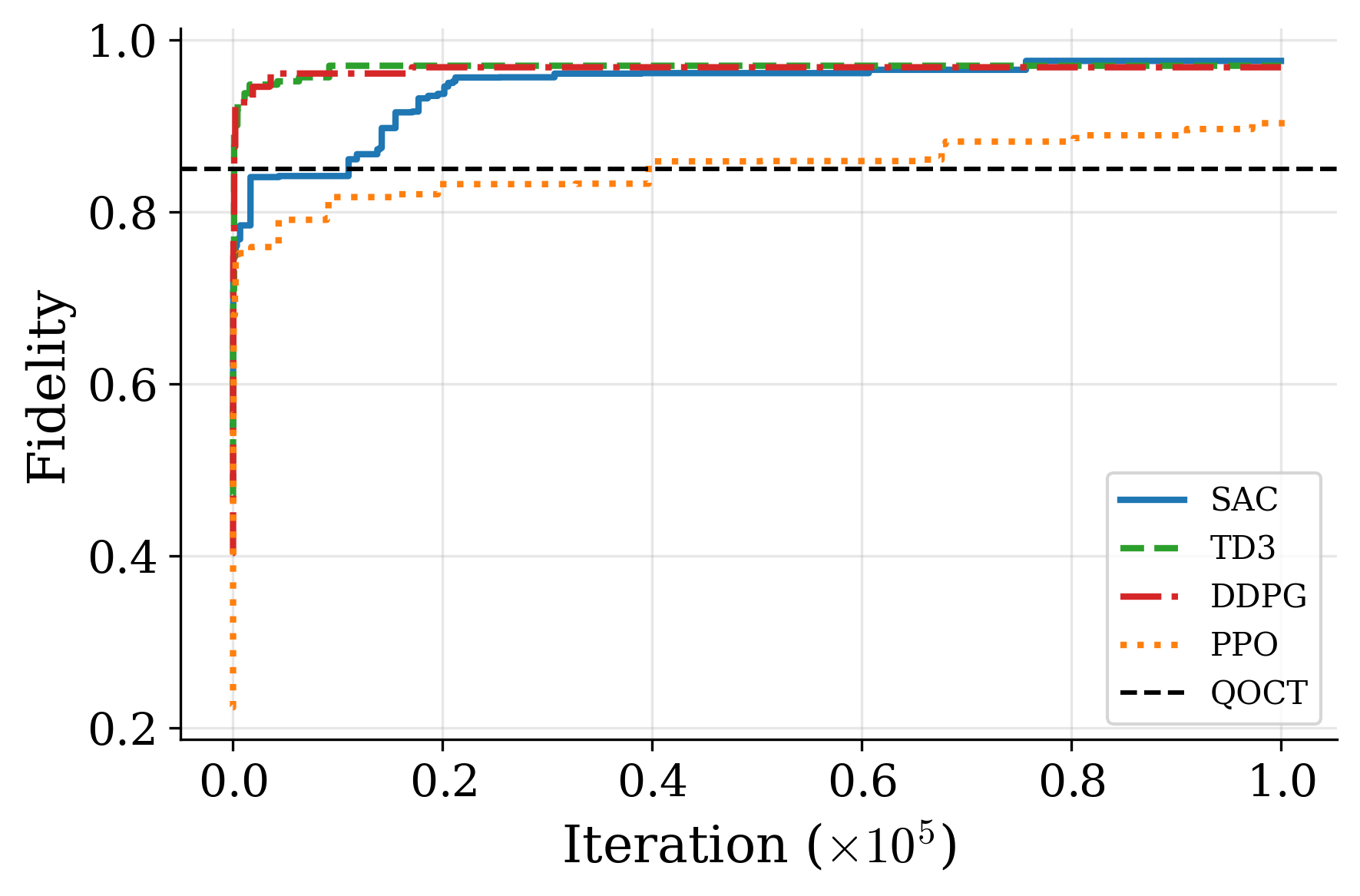}
    \caption{%Top: Gaussian distributions of static parameter offsets used
    %for training and evaluation. Bottom: t
    Training curves for
    $U_{\mathrm{alt}}$, showing rapid convergence of TD3 and DDPG, stable
    SAC performance, and slower PPO learning.}
    \label{fig:alt_noise_and_learning}
\end{figure}

\begin{figure}[h]
    %\centering
    \includegraphics[width=0.8\columnwidth]{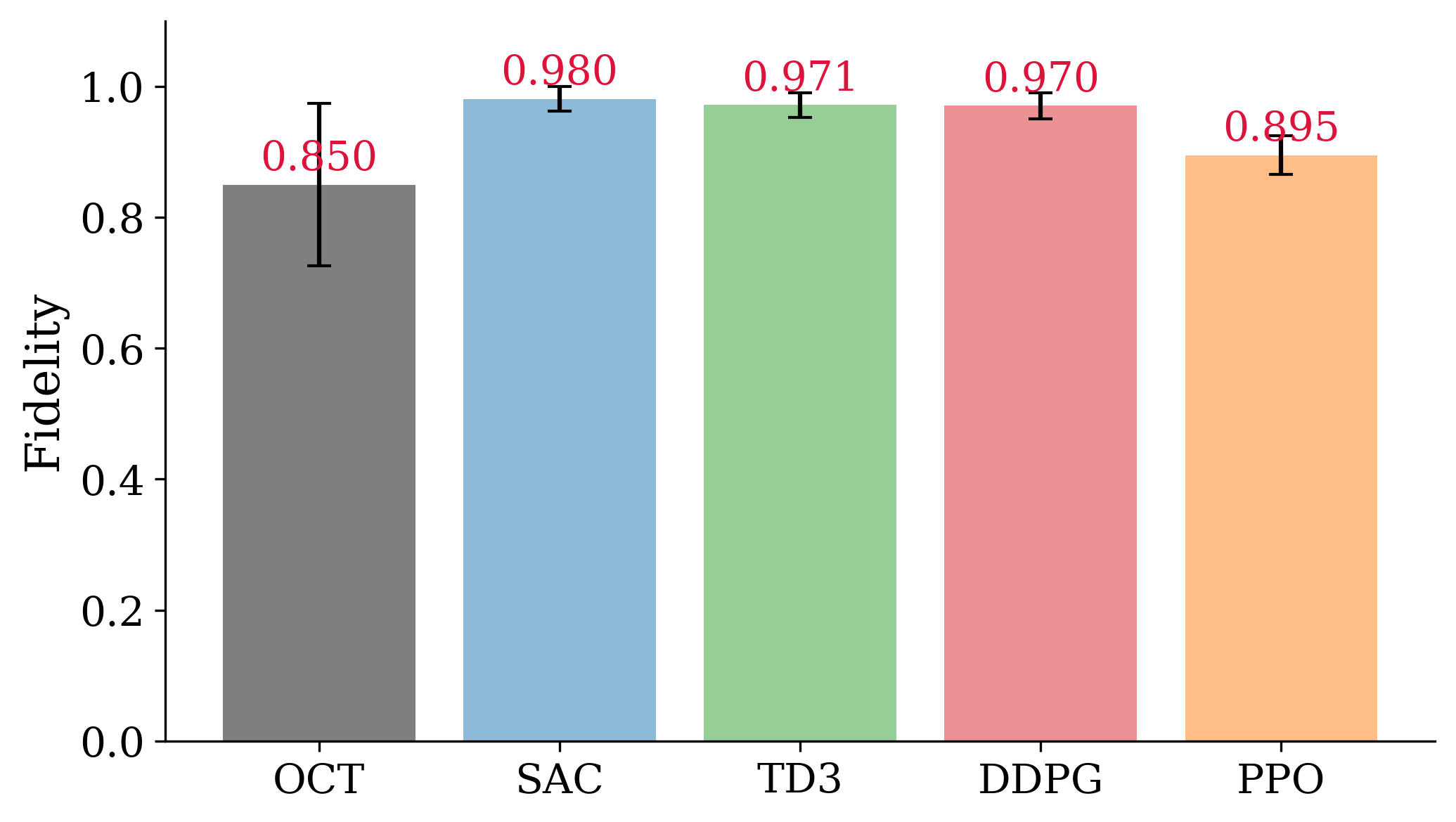}\\[0.5ex]
    \includegraphics[width=0.8\columnwidth]{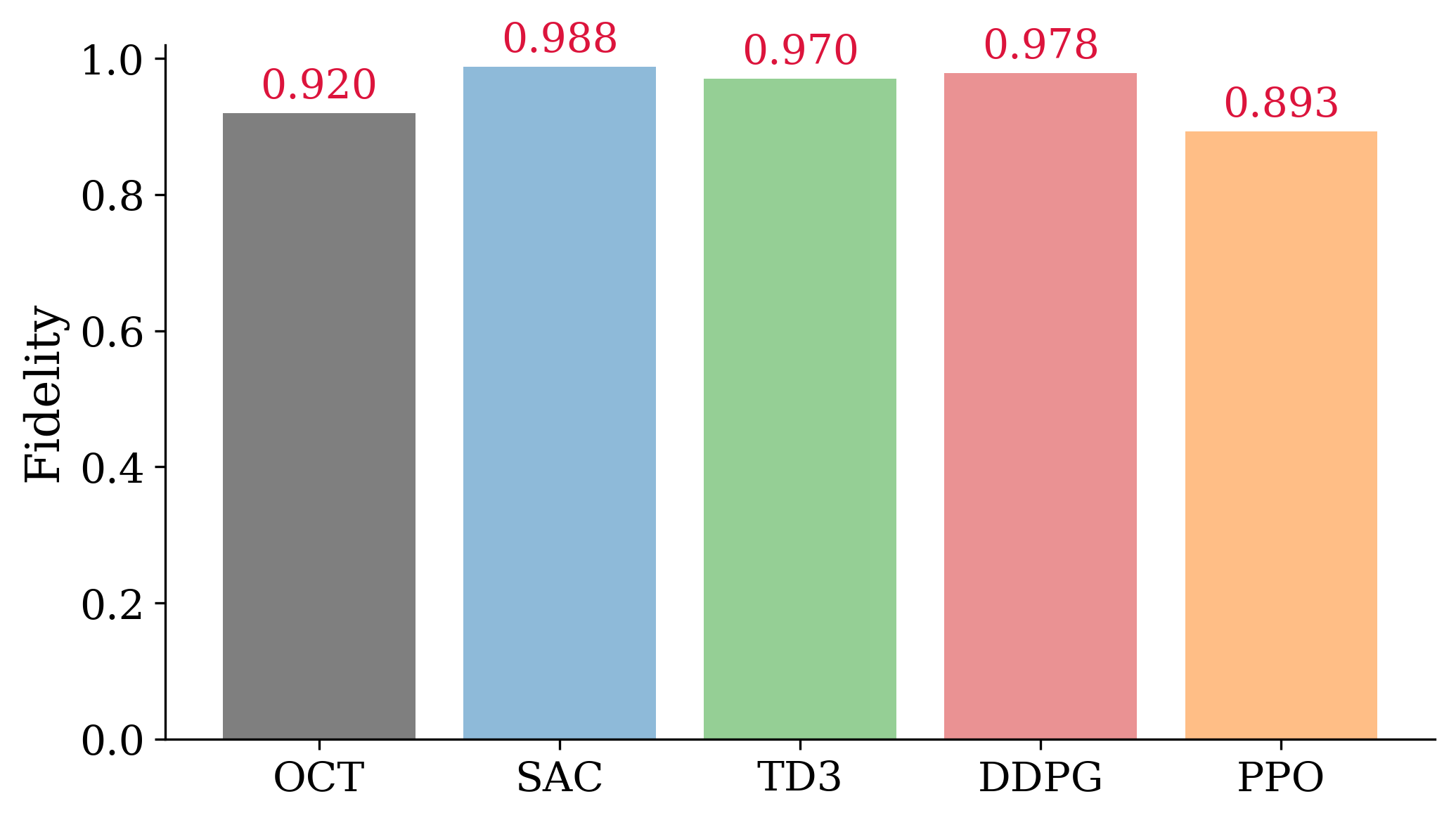}
    \caption{Top: ensemble-averaged fidelity for $U_{\mathrm{alt}}$
    over 100 noisy devices. Bottom: fidelities on a representative fixed noisy device.}
    \label{fig:alt_ensemble_and_single}
\end{figure}

In this Appendix, we briefly validate the generality of the hybrid QOCT+DRL framework using an alternative two-qutrit controlled-phase gate. All simulations use \emph{exactly the same} Hamiltonian parameters, control constraints, noise distributions, training budgets, and DRL architectures as in the main text.
%\subsection{Target gate and nominal device}
The alternative gate is a diagonal two-qutrit unitary acting as
\begin{equation}
U_{\mathrm{alt}} =
\mathrm{diag}(1,1,1,\;
1,1,1,\;
1,1,-1),
\label{eq:U_alt}
\end{equation}
i.e., a $\pi$ phase applied only to the $\ket{22}$ state. This gate provides a minimal entangling benchmark with a simpler phase structure than the $\CZthree$ gate.
Using GRAPE with the same parameters as in the main text
($T=1600$ and $|\epsilon_i|\le 0.3$), QOCT converges reliably to unit fidelity on the nominal Hamiltonian, as shown in Fig.~\ref{fig:alt_QOCT_and_nominal}.  
%\subsection{Residual learning and ensemble robustness}
We next apply the same contextual residual DRL procedure as in the main text. The learning dynamics shown in Fig.~\ref{fig:alt_noise_and_learning} closely mirror those of the $\CZthree$ gate: TD3 and DDPG exhibit the fastest convergence, SAC converges slightly more slowly but
stably, and PPO saturates at a lower fidelity.
The resulting ensemble-averaged fidelities over $M=100$ noisy devices are summarized in Fig.~\ref{fig:alt_ensemble_and_single}. As in the main
text, QOCT alone exhibits both a reduced mean fidelity and a large variance, while all DRL-enhanced strategies significantly improve the average performance and suppress device-to-device fluctuations.

\begin{figure}[h]
    %\centering
    \includegraphics[width=0.8\columnwidth]{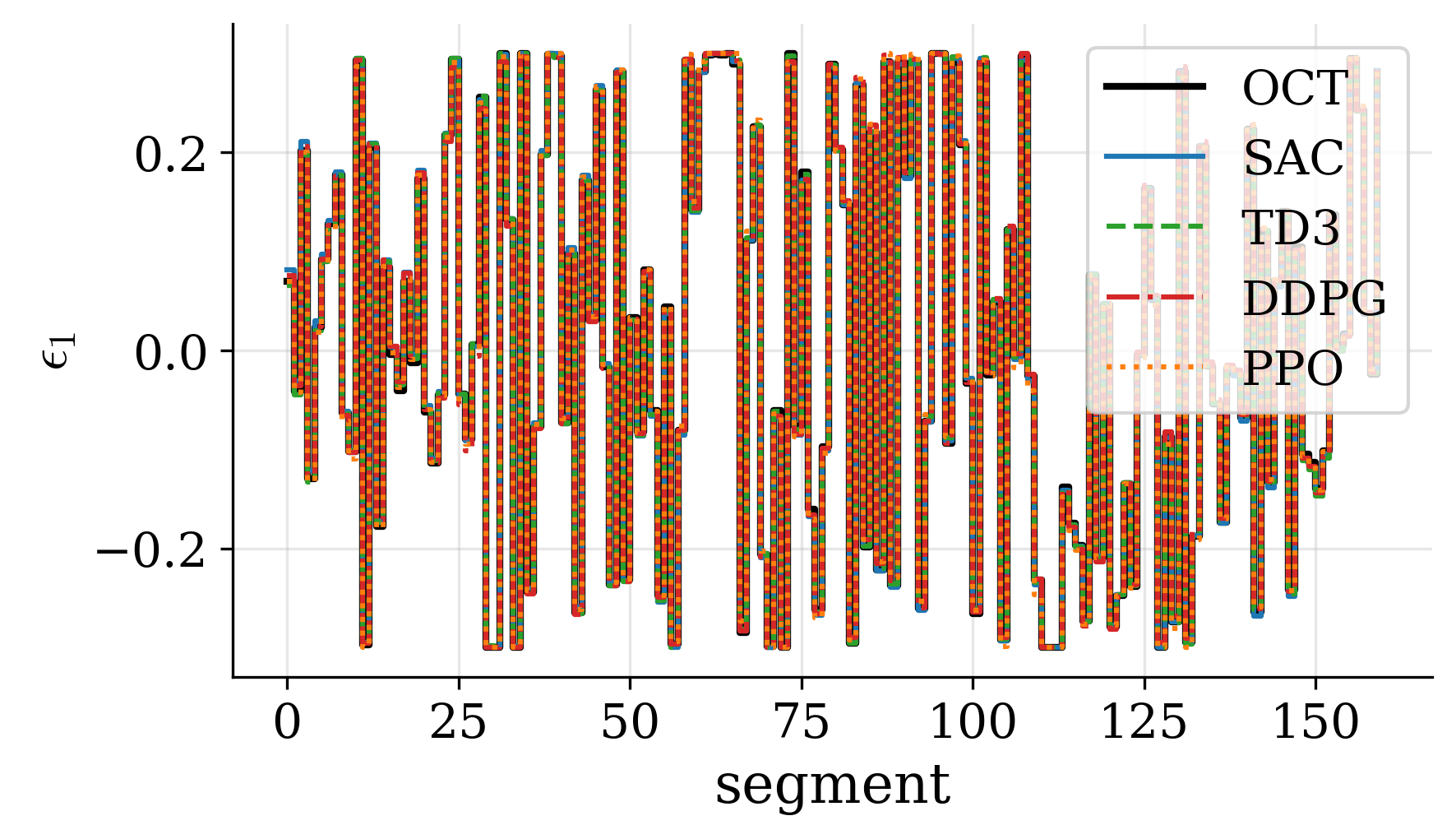}
    \caption{Optimized control drive $\epsilon_1(t)$ for $U_{\mathrm{alt}}$ on a static-noise device.}
    \label{fig:alt_pulses}
\end{figure}

Finally, Fig.~\ref{fig:alt_pulses} shows the optimized drive
$\epsilon_1(t)$ on a representative noisy device. As for the $\CZthree$ gate, the DRL-enhanced pulses remain extremely close to the QOCT baseline, differing only by small, structured corrections distributed over the gate duration.
%The close agreement between the results for $U_{\mathrm{alt}}$ and for the $\CZthree$ gate demonstrates that the hybrid QOCT+DRL approach is insensitive to the detailed phase structure of the target unitary. QOCT provides near-optimal nominal control, while DRL consistently improves robustness by learning low-dimensional residual corrections that compensate for static parameter mismatch.
% ======================================================
% References
% ======================================================

\bibliographystyle{apsrev4-2}

\end{document}